\documentstyle[12pt,epsfig]{article}

\catcode`\@=11
\def\@citex[#1]#2{%
\if@filesw \immediate \write \@auxout {\string \citation {#2}}\fi
\@tempcntb\m@ne \let\@h@ld\relax \def\@citea{}%
\@cite{%
  \@for \@citeb:=#2\do {%
    \@ifundefined {b@\@citeb}%
      {\@h@ld\@citea\@tempcntb\m@ne{\bf ?}%
      \@warning {Citation `\@citeb ' on page \thepage \space undefined}}%
      {\@tempcnta\@tempcntb \advance\@tempcnta\@ne%
      \@tempcntb\number\csname b@\@citeb \endcsname \relax%
      \ifnum\@tempcnta=\@tempcntb 
        \ifx\@h@ld\relax%
          \edef \@h@ld{\@citea\csname b@\@citeb\endcsname}%
        \else%
          \edef\@h@ld{\ifmmode{-}\else--\fi\csname b@\@citeb\endcsname}%
        \fi%
      \else
        \@h@ld\@citea\csname b@\@citeb \endcsname%
        \let\@h@ld\relax%
      \fi}%
    \def\@citea{,\penalty\@highpenalty\,}%
  }\@h@ld
}{#1}}

\def\@citeb#1#2{{[#1]\if@tempswa , #2\fi}}
\def\@citeu#1#2{{$^{#1}$\if@tempswa , #2\fi }}
\def\@citep#1#2{{#1\if@tempswa , #2\fi}}

\def\bcites{         
        \catcode`\@=11
        \let\@cite=\@citeb
        \catcode`\@=12
}

\def\upcites{         
        \catcode`\@=11
        \let\@cite=\@citeu
        \catcode`\@=12
}

\def\plaincites{      
        \catcode`\@=11
        \let\@cite=\@citep
        \catcode`\@=12
}

\newcount\hour
\newcount\minute
\newtoks\amorpm
\hour=\time\divide\hour by 60
\minute=\time{\multiply\hour by 60 \global\advance\minute by-\hour}
\edef\standardtime{{\ifnum\hour<12 \global\amorpm={am}%
        \else\global\amorpm={pm}\advance\hour by-12 \fi
        \ifnum\hour=0 \hour=12 \fi
        \number\hour:\ifnum\minute<10 0\fi\number\minute\the\amorpm}}
\edef\militarytime{\number\hour:\ifnum\minute<10 0\fi\number\minute}

\def\draftlabel#1{{\@bsphack\if@filesw {\let\thepage\relax
   \xdef\@gtempa{\write\@auxout{\string
      \newlabel{#1}{{\@currentlabel}{\thepage}}}}}\@gtempa
   \if@nobreak \ifvmode\nobreak\fi\fi\fi\@esphack}
        \gdef\@eqnlabel{#1}}
\def\@eqnlabel{}
\def\@vacuum{}
\def\marginnote#1{}
\def\draftmarginnote#1{\marginpar{\raggedright\scriptsize\tt#1}}
\def\draft{
        \pagestyle{plain}
        \overfullrule=2pt
        \oddsidemargin -.5truein
        \def\@oddhead{\sl \phantom{\today\quad\militarytime} \hfil
        \smash{\Large\sl DRAFT} \hfil \today\quad\militarytime}
        \let\@evenhead\@oddhead
        \let\label=\draftlabel
        \let\marginnote=\draftmarginnote
        \def\ps@empty{\let\@mkboth\@gobbletwo
        \def\@oddfoot{\hfil \smash{\Large\sl DRAFT} \hfil}
        \let\@evenfoot\@oddhead}
        \def\@eqnnum{(\theequation)\rlap{\kern\marginparsep\tt\@eqnlabel}%
        \global\let\@eqnlabel\@vacuum}  }

\def\blackfonts{
        \font\blackboard=msbm10 scaled\magstep1
        \font\blackboards=msbm8
        \font\blackboardss=msbm6
}

\def\nblack{            
        \def\ZZ{{Z \n{10} Z}}
        \def\NN{{N \n{14} N}}
        \def\RR{{R \n{11} R}}
        \def\QQ{{Q \n{12} Q}}
}

\def\prep{         
        \catcode`\@=11
        \input art10.sty
        \catcode`\@=12
        
        \let\small\null
        \def\blackfonts{
                \font\blackboard=msbm10
                \font\blackboards=msbm7
                \font\blackboardss=msbm5
        }
        \let\sl\it
        \twocolumn
        \sloppy
        \voffset=-2.54truecm
        \hoffset=-2.54truecm
        \flushbottom
        \parindent 1em
        \leftmargini 2em
        \leftmarginv .5em
        \leftmarginvi .5em
        \marginparwidth 48pt
        \marginparsep 10pt
        \setlength{\columnsep}{2truecm}
        \setlength{\textwidth}{25.4truecm}
        \setlength{\textheight}{17truecm}
        \baselineskip=16pt
        \oddsidemargin .18truein
        \evensidemargin .17truein
}

\def\eqalign#1{\null\,\vcenter{\openup\jot\m@th
  \ialign{\strut\hfil$\displaystyle{##}$&$\displaystyle{{}##}$\hfil
      \crcr#1\crcr}}\,}
\def\eqalignno#1{\displ@y \tabskip\centering
  \halign to\displaywidth{\hfil$\@lign\displaystyle{##}$\tabskip\z@skip
    &$\@lign\displaystyle{{}##}$\hfil\tabskip\centering
    &\llap{$\@lign##$}\tabskip\z@skip\crcr
    #1\crcr}}

\def\section{\@startsection {section}{1}{\z@}{3.ex plus 1ex minus
 .2ex}{2.ex plus .2ex}{\large\bf}}
\def\subsection{\@startsection{subsection}{2}{\z@}{2.75ex plus 1ex minus
 .2ex}{1.5ex plus .2ex}{\bf}}        

\def\appendix{{\newpage\section*{Appendix}}\let\appendix\section%
        {\setcounter{section}{0}
        \gdef\thesection{\Alph{section}}}\section}

\def\abstract{\if@twocolumn
\section*{Abstract}
\else 
\begin{center}
{\bf Abstract\vspace{-.5em}\vspace{0pt}}
\end{center}
\quotation
\fi}

\catcode`\@=12

\newcommand{\beq}{\begin{equation}}
\newcommand{\eeq}{\end{equation}}
\newcommand{\beqa}{\begin{eqnarray}}
\newcommand{\eeqa}{\end{eqnarray}}

\newcommand{\Z}{{\bf Z}}

\newcommand{\R}{{\bf R}}
\newcommand{\C}{{\bf C}}
\newcommand{\e}{\,{\rm e}}

\newcommand{\dd}{{\rm d}}

\def\noj#1,#2,{{\bf #1} (19#2)\ }
\def\jou#1,#2,#3,{{\sl #1\/ }{\bf #2} (19#3)\ }
\def\ann#1,#2,{{\sl Ann.\ Physics\/ }{\bf #1} (19#2)\ }
\def\cmp#1,#2,{{\sl Comm.\ Math.\ Phys.\/ }{\bf #1} (19#2)\ }
\def\ma#1,#2,{{\sl Math.\ Ann.\/ }{\bf #1} (19#2)\ }
\def\ng#1,#2,{{\sl Nagoya.\ Math.\ J.\/ }{\bf #1} (19#2)\ }
\def\jd#1,#2,{{\sl J.\ Diff.\ Geom.\/ }{\bf #1} (19#2)\ }
\def\invm#1,#2,{{\sl Invent.\ Math.\/ }{\bf #1} (19#2)\ }
\def\cq#1,#2,{{\sl Class.\ Quantum Grav.\/ }{\bf #1} (19#2)\ }
\def\cqg#1,#2,{{\sl Class.\ Quantum Grav.\/ }{\bf #1} (19#2)\ }
\def\ijmp#1,#2,{{\sl Int.\ J.\ Mod.\ Phys.\/ }{\bf A#1} (19#2)\ }
\def\jmphy#1,#2,{{\sl J.\ Geom.\ Phys.\/ }{\bf #1} (19#2)\ }
\def\jams#1,#2,{{\sl J.\ Amer.\ Math.\ Soc.\/ }{\bf #1} (19#2)\ }
\def\grg#1,#2,{{\sl Gen.\ Rel.\ Grav.\/ }{\bf #1} (19#2)\ }
\def\mpl#1,#2,{{\sl Mod.\ Phys.\ Lett.\/ }{\bf A#1} (19#2)\ }
\def\nc#1,#2,{{\sl Nuovo Cim.\/ }{\bf #1} (19#2)\ }
\def\np#1,#2,{{\sl Nucl.\ Phys.\/ }{\bf B#1} (19#2)\ }
\def\pl#1,#2,{{\sl Phys.\ Lett.\/ }{\bf #1B} (19#2)\ }
\def\pla#1,#2,{{\sl Phys.\ Lett.\/ }{\bf #1A} (19#2)\ }
\def\pr#1,#2,{{\sl Phys.\ Rev.\/ }{\bf #1} (19#2)\ }
\def\prd#1,#2,{{\sl Phys.\ Rev.\/ }{\bf D#1} (19#2)\ }
\def\prl#1,#2,{{\sl Phys.\ Rev.\ Lett.\/ }{\bf #1} (19#2)\ }
\def\prp#1,#2,{{\sl Phys.\ Rept.\/ }{\bf #1C} (19#2)\ }
\def\ptp#1,#2,{{\sl Prog.\ Theor.\ Phys.\/ }{\bf #1} (19#2)\ }
\def\ptpsup#1,#2,{{\sl Prog.\ Theor.\ Phys.\/ Suppl.\/ }{\bf #1} (19#2)\ }
\def\rmp#1,#2,{{\sl Rev.\ Mod.\ Phys.\/ }{\bf #1} (19#2)\ }
\def\yadfiz#1,#2,#3[#4,#5]{{\sl Yad.\ Fiz.\/ }{\bf #1} (19#2) #3%
\ [{\sl Sov.\ J.\ Nucl.\ Phys.\/ }{\bf #4} (19#2) #5]}
\def\zh#1,#2,#3[#4,#5]{{\sl Zh.\ Exp.\ Theor.\ Fiz.\/ }{\bf #1} (19#2) #3%
\ [{\sl Sov.\ Phys.\ JETP\/ }{\bf #4} (19#2) #5]}

\def\beq{\begin{equation}}
\def\eeq{\end{equation}}
\def\beqar{\begin{eqnarray}}
\def\eeqar{\end{eqnarray}}

\newcommand{\be}{\begin{equation}}
\newcommand{\ee}{\end{equation}}
\newcommand{\bea}{\begin{eqnarray}}
\newcommand{\eea}{\end{eqnarray}}

\def\nfrac#1#2{{\displaystyle{\vphantom1\smash{\lower.5ex\hbox{\small$#1$}}%
        \over\vphantom1\smash{\raise.25ex\hbox{\small$#2$}}}}}

\def\n#1{\mskip-#1mu}

\def\to{\rightarrow}

\def\lae{\mathrel{\mathop{\smash{\lower .5 ex \hbox{$\stackrel<\sim$}}}}}
\def\lae{\mathrel{\mathop{\smash{\lower .5 ex \hbox{$\stackrel>\sim$}}}}}

\def\l:{\mathopen{:}\,}
\def\r:{\,\mathclose{:}}

\catcode`\@=11
\def\theequation{\arabic{equation}}
\catcode`\@=12

\nblack
\bcites

\nblack

\catcode`\@=11
\def\theequation{\thesection.\arabic{equation}}
\@addtoreset{equation}{section}
\@addtoreset{footnote}{section}
\@addtoreset{footnote}{subsection}
\catcode`\@=12

\newcommand{\beqn}{\begin{equation}}
\newcommand{\eeqn}{\end{equation}}
\newcommand{\beqnarray}{\begin{eqnarray}}
\newcommand{\eeqnarray}{\end{eqnarray}}
\newcommand {\bear} [1] {\begin {array} {#1}}
\newcommand {\ear} {\end {array}}

\newcommand{\CP}{{\bf C}{\rm P}}

\newcommand {\beqarn} {\begin{eqnarray*}}
\newcommand {\eeqarn} {\end{eqnarray*}}

\newcommand{\comment}[1]{}

\def\bbbz{{\sf Z\!\!\!Z}}
\def\sl2z{SL(2,\bbbz)}

\newcommand{\nn}{\nonumber}

\newcommand{\opsi}{\overline{\psi}}
\newcommand{\oQ}{\overline{Q}}

\newcommand{\oep}{\overline{\epsilon}}
\newcommand{\eps}{\epsilon}

\newcommand{\btheta}{\overline{\theta}}
\newcommand{\cQ}{{\cal Q}}
\newcommand{\bepsilon}{\overline{\epsilon}}
\newcommand{\bPhi}{\overline{\Phi}}
\newcommand{\bphi}{\overline{\phi}}
\newcommand{\bpsi}{\overline{\psi}}

\newcommand{\bi}{\overline{\imath}}
\newcommand{\bj}{\overline{\jmath}}
\newcommand{\bz}{\overline{z}}
\newcommand{\bchi}{\overline{\chi}}
\newcommand{\blambda}{\overline{\lambda}}
\newcommand{\bsigma}{\overline{\sigma}}
\newcommand{\bareta}{\overline{\eta}}
\newcommand{\lrd}{\overleftarrow{\partial}\!\!\!\!\!
\overrightarrow{\vbox to 8.35pt{}}}
\newcommand{\lrD}{\overleftarrow{D}\!\!\!\!\!
\overrightarrow{\vbox to 8.2pt{}}}
\newcommand{\bD}{\overline{D}}
\newcommand{\bcQ}{\overline{\cal Q}}

\begin{document}

\begin{titlepage}

\begin{center}

\today\hfill
HUTP-00/A051\\
\hfill hep-th/0012179

\vskip 1.5 cm
{\large \bf Linear Models of Supersymmetric D-Branes}
\vskip 1 cm 
{Kentaro Hori}\\
\vskip 0.5cm
{\it Jefferson Physical Laboratory,
Harvard University\\
Cambridge, MA 02138, U.S.A.}\\

\end{center}

\vskip 0.5 cm
\begin{abstract}

We construct a class of supersymmetric boundary interactions
in ${\cal N}=2$ field theories on the half-space,
which depend on parameters that are not at all renormalized
or not renormalized in perturbation theory beyond one-loop.
This can be used to study
D-branes wrapped on a certain class of Lagrangian submanifolds
as well as holomorphic cycles.
The construction of holomorphic D-branes
is in close relationship with
the background independent open string field theory approach to
brane/anti-brane systems.
As an application, mirror pairs of Lagrangian and holomorphic
D-branes are identified.
The mirror pairs are studied by twisting to open topological field theories.

\end{abstract}

\end{titlepage}

\newpage

\section{Introduction}\label{sec:intro}

Let us consider the $1+1$ dimensional $U(1)$ gauge theory
of charge $1$ complex scalar fields $\phi_1,\ldots,\phi_N$
with the following action
\beq
S={1\over 2\pi}\int
\dd^2x\left\{
-\sum_{i=1}^N|D_{\mu}\phi_i|^2
+D\left(\sum_{i=1}^N|\phi_i|^2-r\right)
+\theta v_{01}
\right\}.
\label{LSM}
\eeq
Here $D$ is an auxiliary field and $v_{01}=\partial_0 v_1-\partial_1 v_0$
is the fieldstrength of the $U(1)$ gauge potential $v_{\mu}$.
Eliminating the auxiliary field $D$ we obtain the constraint
\beq
\sum_{i=1}^N|\phi_i|^2=r.
\eeq
One can also solve for $v_{\mu}$ and, after modding out by the $U(1)$
gauge group action, we obtain the non-linear sigma model whose target space
is the complex projective space $\CP^{N-1}$.
The Theta term $\theta v_{01}$ becomes the $B$-field term
which is non-zero for a topologically non-trivial
field configuration. 
The above gauge theory is called the linear sigma model for $\CP^{N-1}$.

Linear sigma models have played important roles in understanding
several dynamical aspects of quantum field theories.
In recent years, ${\cal N}=2$ supersymmetric
linear sigma models in $1+1$ dimensions has been used effectively
to understand some of the key aspects of supersymmetric
non-linear sigam models and related models.
An advantage in the construction \cite{phases}
is that the parameters that are not renormalized or
renormalized only at the one-loop level are explicitly identified
and simply realized.
For instance, in the supersymmetric generalization of the model
(\ref{LSM}), the complex combination
\beq
t=r-i\theta
\eeq
appears in a twisted superpotential and it is manifest that it
is renormalized only at the one-loop level.
Sigma model on a hypersurface of $\CP^{N-1}$ can also be realized as
a gauge theory and the complex structure parameters enter into
the superpotential term, which is not renormalized and is decoupled from
the Kahler class parameters.
One can make a precise statement
on the theory only after such an identification of parameters
is made.
The proof of mirror symmetry in \cite{HV} makes use of this
advantage of the construction.

It is natural to ask whether there is a similar construction
for theories formulated on a worldsheet with boundaries.
The boundary conditions and interactions are the new ingredient. 
Such theories are the relevant models in defining and analyzing
open strings or D-branes in string theory.
Recently, great progress has been made in understanding
several aspects of supersymmetric D-branes \cite{BBS} in
Calabi-Yau and non-geometric compactification.
In particular, a lot has been understood and clarified from
the worldsheet points of view
\cite{SYZ,OOY,DGep,BDLR,DGomi,GJS0,DGep2,KKLM,Sugi,HIV,lerche,GJS,Kapustin,DCat,HaNa,AV}.
(Earlier relevant works are in \cite{WCS,nick}.
See also \cite{DFR,Iqbal,DDoug} for some studies
from other approaches.)
However, we still lack
a worldsheet formulation
with boundary interactions depending on parameters
that are not renormalized or simply renormalized.
We expect to learn a lot more by having
such a formulation.\footnote{A similar
point was also made in \cite{shamit}.}
The aim of this paper is to construct
such a formulation.

As a basic example, let us consider a D1-brane located
at the circle $|z|^2=c$ in the complex plane $\C$.
The worldsheet theory is described in terms of
a complex scalar field $\phi$ with the action
$S=-{1\over 2\pi}\int \dd^2x |\partial_{\mu}\phi|^2$.
We impose
the following condition
at the worldsheet boundary
\beq
\begin{array}{cl}
\mbox{(D)}&~~
|\phi|^2=c,\\[0.1cm]
\mbox{(N)}&~~
\partial_n\arg(\phi)=0,
\end{array}
\label{bcon}
\eeq
where $\partial_n$ is the normal derivative.
One can also consider adding the boundary term in the action
\beq
S_a={a\over 2\pi}\int\limits_{\partial\Sigma}
 \dd\arg(\phi),
\label{Sa}
\eeq
which is non-zero for a topologically non-trivial configuration.
The boundary condition breaks the scale invariance of the
bulk theory and the parameter $c$ is renormalized
at the one loop level as
$c(\mu)=c(\mu')+\log(\mu/\mu')$ \cite{Leigh}
(as reviewed in Secion~\ref{subsec:ra}).
In the supersymmetric generalization of this model,
the axial $U(1)$ R-symmetry is anomalous
or the axial rotation shifts the parameter $a$.
This suggests that the parameters $c$ and $a$ are superpartners
of each other and it is natural to consider
the complex combination
\beq
s=c-ia.
\label{defs}
\eeq
We will construct a boundary interaction
that induces (\ref{Sa}) and also yields the boundary condition
(\ref{bcon}) in such a way that the parameter $s$
enters in some kind of superpotential term on the boundary.
The basic idea is to introduce auxiliary degrees of freedom,
just like $D$ and $v_{\mu}$ in the $\CP^{N-1}$ model, but living only
on the worldsheet boundary.
When applied to more general bulk theories,
this will lead to the construction of supersymmetric boundary theories
for D-branes wrapped on a class of Lagrangian submanifolds
of the target space.
The parameters will be protected from (higher) loop corrections
although they can recive non-perturbative corrections.

Another type of supersymmetric boundary theories correspond to
D-branes wrapped on holomorphic cycles, or those supporting
holomorphic vector bundles.
We will construct simple models that realizes a certain class
of such D-branes, where the parameters characterizing the D-branes
enter into a boundary superpotential term.
The construction is closely related to
the worldsheet approach to the problem of the tachyon condensation
of unstable D-brane systems \cite{KHM,KMM2,Justin}.
(The latter reference was considered in the context of
the background independent open string field theory
\cite{GShat,KMM1,BIOSFT}.)
In this construction, we will also have boundary degrees of freedom
as the essential ingredients.
Our construction gives rise to a simple and explicit 
realization of a certain class of complexes of sheaves
that has been argued to be the basic elements in
supersymmetric D-branes \cite{DCat}.

The rest of the paper is organized as follows.
In Section~\ref{sec:BSS}, we introduce superspace formalism on the
worldsheet with boundary.
In Section~\ref{sec:AD}, the supersymmetric version of the D1-brane
discussed above is studied and it is shown that (\ref{defs}) is
the chiral parameter. We then
construct a ``linear model'' for this basic example.
In Section~\ref{sec:LinA},
we apply the construction to certain supersymmetric gauge theories.
This leads to the worldsheet definition of a certain class of
A-type Lagrangian D-branes in toric manifolds.
We also find the mirror description of such D-branes.
In Section~\ref{sec:BTachy} we construct the boundary interactions
corresponding to B-type holomorphic
D-branes. We start with space-filling brane/anti-brane system
and find the condition
for ${\cal N}=2$ worldsheet supersymmetry.
In Section~\ref{sec:BLG}, we study B-type D-branes in LG models.
We especially study in detail the properties of D0-branes
which are the mirrors of toroidal 
Lagrangian D-branes.

\section{${\cal N}=2$ Boundary Superspace}\label{sec:BSS}

In this paper,
we consider $1+1$ dimensional field theories
with $(2,2)$ supersymmetry in the bulk,
a half of which is preserved by
the boundary conditions or boundary interactions.
In order to make the supersymmetry structure manifest
and to identify parameters that are not renormalized or
do not receive perturbative renormalization beyond one loop,
it is convenient to introduce the superspace formalism
on the worldsheet with boundaries.

\subsection{$(2,2)$ Superspace}

To fix the notation, we briefly describe here
the superspace formalism for $(2,2)$ supersymmetry in the bulk.

The $(2,2)$ superspace has four fermionic coordinates
$\theta^+,\,\theta^-,\,\btheta^+,\,\btheta^-$,
in addition to the bosonic coordinates $x^0, \,x^1$.
The supersymmetry transformations
are generated by the following differential operators on the superspace,
\bea
{\cal Q}_{\pm}&=&\frac{\partial}{\partial \theta^{\pm}}
+i \overline{\theta}^{\pm}\,\partial_{\pm},
\label{Qpm}\\
\overline{{\cal Q}}_{\pm}
&=&
-\frac{\partial}{\partial \overline{\theta}^{\pm}}
-i \theta^{\pm}\,\partial_{\pm},
\label{oQpm}
\eea
where $\partial_{\pm}$ are differentiations by $x^{\pm}=x^0\pm x^1$;
\beq
\partial_{\pm}={\partial\over \partial x^{\pm}}=
{1\over 2}\left({\partial\over \partial x^0}
\pm{\partial\over \partial x^1}\right).
\eeq
These differential operators obey the anti-commutation relations
$\{ {\cal Q}_{\pm},\overline{{\cal Q}}_{\pm}\} =-2 i \partial_{\pm}$.
We introduce another set of differential operators
\bea
D_{\pm}&=&\frac{\partial}{\partial \theta^{\pm}}
-i \overline{\theta}^{\pm}\,\partial_{\pm}, \\
\overline{D}_{\pm}&=&-\frac{\partial}{\partial \overline{\theta}^{\pm}}
+i \theta^{\pm}\,\partial_{\pm},
\eea
 which anti-commute
with ${\cal Q}_{\pm}$ 
and $\overline{{\cal Q}}_{\pm}$.
These obey the similar anti-commutation relations
$\{D_{\pm},\overline{D}_{\pm}\}=2i\partial_{\pm}$.
{\it Vector R-rotation} and {\it axial R-rotation}
of a superfield are defined by
\bea
\e^{i\alpha F_{V}}:{\cal F}(x^{\mu},\theta^{\pm},\btheta^{\pm})\mapsto
\e^{i \alpha q_{V}}
{\cal F}(x^{\mu},\e^{-i\alpha}\theta^{\pm},
\e^{i\alpha}\overline{\theta}^{\pm})\\
\e^{i\beta F_{A}}:{\cal F}(x^{\mu},\theta^{\pm},\btheta^{\pm})\mapsto
\e^{i \beta q_{A}}{\cal F}(x^{\mu},\e^{\mp i\beta}\theta^{\pm},
\e^{\pm i\beta}\overline{\theta}^{\pm}),
\eea
where $q_V$ and $q_A$ are numbers called vector R-charge and
axial R-charge of ${\cal F}$.
A {\it chiral superfield} $\Phi$ is a superfield
which satisfies the equations,
$\overline{D}_{\pm}\Phi =0$.
It has the follwing expansion
($y^{\pm}:=x^{\pm}-i\theta^{\pm}\btheta^{\pm}$)
\beq
\Phi(x^{\mu},\theta^{\pm},\btheta^{\pm})
=\phi(y)+\theta^{\alpha}\psi_{\alpha}(y)
+\theta^+\theta^-F(y).
\eeq
A {\it twisted chiral superfield} $\widetilde{\Phi}$
is a superfield which satisfies
$\overline{D}_+\widetilde{\Phi}=D_-\widetilde{\Phi}=0$.
It has the following expansion
($\widetilde{y}^{\pm}:=x^{\pm}\mp i\theta^{\pm}\btheta^{\pm}$)
\beq
\widetilde{\Phi}(x^{\mu},\theta^{\pm},\btheta^{\pm})
=\widetilde{\phi}(\widetilde{y})
+\theta^+\bchi_+(\widetilde{y})
+\btheta^-\chi_-(\widetilde{y})
+\theta^+\btheta^-E(\widetilde{y}).
\eeq

There are three kind of action functionals of supefields
which are invariant under the
supersymmetry transformation
$\delta
=\epsilon_+\cQ_--\epsilon_-\cQ_+-\bepsilon_+\overline{\cQ}_-
+\bepsilon_-\overline{\cQ}_+$.
One is the D-term integral
\be
\int \dd^{2}x \,\dd^{4}\theta \,
K({\cal F}_i)=\int \dd^2x \,\dd\theta^+\dd\theta^-\dd\btheta^-\dd\btheta^+\,
K({\cal F}_i),
\label{Dterm}
\ee
where $K(-)$ is an arbitrary differentiable function
of arbitrary superfields ${\cal F}_i$.
The next is the F-term integral
\be
\int \dd^{2}x \dd^{2}\theta\, W(\Phi_i)
=\int \dd^2x\, \dd\theta^- \dd\theta^+\,
W(\Phi_i)\Bigr|_{\btheta^{\pm}=0},
\label{Fterm}
\ee 
where $W(\Phi_i)$ is a holomorphic function of chiral superfields
$\Phi_i$ which is called a superpotential.
The third is the twisted F-term integral
\beq
\int \dd^2x\,\dd^2\widetilde{\theta}\,\widetilde{W}(\widetilde{\Phi}_i)
=\int \dd^2x\,\dd\btheta^-\dd\theta^+\,\widetilde{W}(\widetilde{\Phi}_i)
\Bigr|_{\btheta^+=\theta^-=0},
\label{tFterm}
\eeq
where $\widetilde{W}(\widetilde{\Phi}_i)$ is a holomorphic function of
twisted chiral superfields $\widetilde{\Phi}_i$
which is called a twisted superpotential.

Chiral superfields cannot enter into twisted F-term and
twisted chiral superfields cannot enter into F-term.
Using the argument of Seiberg \cite{seiberg},
one can also show that the parameters that enters into
the superpotential (resp. twisted superpotential)
at a high energy scale cannot enter
into the twisted superpotential (resp. superpotential)
in the effective action at a lower energy.

\subsection{Superspace Boundaries}

Let us put a $(2,2)$ supersymmetric field theory
on the ``left half plane''
\beq
\Sigma=\R\times (-\infty,0],
\eeq
where $\R$ and $(-\infty,0]$
are parametrized by the time and spacial coordinates
$-\infty<x^0<+\infty$, $-\infty<x^1\leq 0$
respectively. The left half plane $\Sigma$ has its boundary
$\partial\Sigma$ at
\beq
x^1=0.
\label{bosb}
\eeq
We suppose that there is an analogous ``boundary''
in the fermionic coordinates
$\theta^{\pm},\btheta^{\pm}$ as well.
We now determine what kind of boundary is possible.\footnote{A part of
the construction in this section has been made
by E. Martinec \cite{emil}.
In particular, (\ref{Ab}) and (\ref{Bb}) was stated in \cite{emil}.
 We thank him for letting us know about it.
Boundary superspace was also considered in \cite{ItoMox} for
${\cal N}=1$ supersymmetry.}

To define a consistent theory, one must impose some boundary condition
on the fields. The boundary condition
usually relates the left moving modes and the right
moving modes. In particular, the left moving and the right moving fermions
are related to each other.
This suggests that the boundary relates the fermionic coordinates
$\theta^+, \btheta^+$ and the other coordinates
$\theta^-, \btheta^-$.
There are essentially two ways to relate them;
\medskip
\beqa
\mbox{(A)}&&\theta^++\e^{i\alpha}\btheta^-=0,~~~
\btheta^++\e^{-i\alpha}\theta^-=0,
\label{Ab}\\[0.1cm]
\mbox{(B)}&&\theta^+-\e^{i\beta}\theta^-=0,~~~
\btheta^+-\e^{-i\beta}\btheta^-=0.
\label{Bb}
\eeqa

\noindent
In the above expression, $\e^{i\alpha}$ and $\e^{i\beta}$
are fixed phases. In most of the following discussion,
we shall set these phases trivial, $\e^{i\alpha}=1$,
$\e^{i\beta}=1$.
We call (\ref{bosb}) and (\ref{Ab}) {\it A-boundary}
while (\ref{bosb}) and (\ref{Bb}) will be called {\it B-boundary}.
As we will see below, these superspace boundaries corresponds
to theories in which the following combinations of the supercharges are
conserved;
\beqa
\mbox{(A)}&&\oQ=\oQ_++\e^{i\alpha}Q_-,~~~Q=Q_++\e^{-i\alpha}\oQ_-,
\label{AQ}\\[0.1cm]
\mbox{(B)}&&\oQ=\oQ_++\e^{i\beta}\oQ_-,~~~Q=Q_++\e^{-i\beta}Q_-.
\label{BQ}
\eeqa
We shall call the former A-type supersymmetry and the latter
B-type supersymmetry.

\subsubsection{A-boundary}

We develop here the superspace formalism for A-boundary.
We set $\e^{i\alpha}=1$ but the generalization is straightforward.
We first introduce the fermionic coordinates at the A-boundary
as
\beq
\theta:=\theta^+=-\btheta^-,~~~\btheta:=\btheta^+=-\theta^-.
\eeq
The boundary (\ref{bosb})-(\ref{Ab})
is preserved by $\partial_0$ and the following differential operators
\beqa
&&\bD=\bD_++D_-=-{\partial\over\partial\btheta}+i\theta\partial_0,\\
&&D=D_++\bD_-={\partial\over\partial\theta}-i\btheta\partial_0,\\
&&\bcQ=\bcQ_++\cQ_-=-{\partial\over\partial\btheta}-i\theta\partial_0,
\label{bQA}\\
&&\cQ=\cQ_++\bcQ_-={\partial\over\partial\theta}+i\btheta\partial_0,
\label{QA}
\eeqa
This indeed shows that the supersymmetry preserved by the full theory
should be (\ref{AQ}).
We also note that the boundary (\ref{Ab}) is invariant under the
axial R-rotation whereas the vector R-rotation rotates the phase
$\e^{i\alpha}$ defining the boundary (\ref{Ab}).

The differential operators obey the anti-commutation relations
$\{D,\bD\}=2i\partial_0$, $D^2=\bD^2=0$, and
$\{\cQ,\bcQ\}=-2i\partial_0$, $\cQ^2=\bcQ^2=0$.
A boundary superfield is a function of the boundary coordinates
$x^0$, $\theta$ and $\btheta$,
which transforms under A-type supersymmetry
(with parameter
$\epsilon_+=\bepsilon_-=\epsilon$, $\bepsilon_+=\epsilon_-=\bepsilon$)
by
$\delta=\epsilon \bcQ-\bepsilon \cQ$.
The boundary R-rotation
transforms the boundary superfield ${\cal F}$ as
\beq
{\cal F}(x^0,\theta,\btheta)\mapsto
\e^{iq\gamma}{\cal F}(x^0,\e^{-i\gamma}\theta,\e^{i\gamma}\btheta),
\eeq
where $q$ is called the boundary R-charge of ${\cal F}$.
A boundary superfield ${\mit\Phi}$ is called a {\it boundary chiral
superfield} if it obeys
\beq
\bD\,{\mit \Phi}=0.
\eeq
A boundary chiral superfield has the following Theta expansion
\beq
{\mit\Phi}=\phi(x^0)+\theta\psi(x^0)-i\theta\btheta\partial_0\phi(x^0).
\eeq
We often call a fermionic boundary chiral superfield
a {\it boundary Fermi superfield}.
Given a function $J({\cal F}_i)$ of boundary superfields ${\cal F}_i$,
the following integral is invariant under the supersymmetry
variation $\delta$
\beq
\int \dd x^0\dd\theta\dd\btheta\,J({\cal F}_i).
\label{bD}
\eeq
Also, given a boundary Fermi superfield ${\mit\Psi}$
and a holomorphic function ${\cal V}({\mit\Phi}_i)$
of boundary chiral superfields ${\mit \Phi}_i$, the integral
\beq
\int \dd x^0\dd\theta ~{\mit\Psi}{\cal V}({\mit\Phi})\,\Bigr|_{\btheta=0},
\label{bF}
\eeq
is invariant under the supersymmetry $\delta$.
We call ${\cal V}({\mit\Phi}_i)$ a {\it boundary superpotential}.
The above term is invariant under the boundary R-rotation if
${\mit\Phi}{\cal V}({\mit\Phi}_i)$ has boundary R-charge $1$.
We shall sometimes refer to (\ref{bD}) and (\ref{bF}) as boundary D-term
and boundary F-term respectively.

A bulk superfield restricted to the boundary (\ref{bosb})-(\ref{Ab})
is a boundary superfield.
The boundary R-rotation comes from the axial R-rotation in the bulk.
It is easy to see that
a bulk twisted chiral superfield
restricted on the A-boundary is a boundary chiral superfield.
The boundary superpotential must be a holomorphic function
of the boundary chiral superfields. This strongly constrains
a possible form of quantum corrections, as in \cite{seiberg},
as we will see explicitly in several examples.

\subsubsection{B-boundary}

We briefly repeat the same thing for B-boundary (we again set
$\e^{i\beta}=1$).
The fermionic coordinates at the B-boundary are
\beq
\theta:=\theta^+=\theta^-,~~~\btheta:=\btheta^+=\btheta^-.
\eeq
The following differential operators preserve the
B-boundary (\ref{bosb})-(\ref{Bb});
\beqa
&&\bD=\bD_++\bD_-=-{\partial\over\partial\btheta}+i\theta\partial_0,\\
&&D=D_++D_-={\partial\over\partial\theta}-i\btheta\partial_0,\\
&&\bcQ=\bcQ_++\bcQ_-=-{\partial\over\partial\btheta}-i\theta\partial_0,
\label{bQB}\\
&&\cQ=\cQ_++\cQ_-={\partial\over\partial\theta}+i\btheta\partial_0,
\label{QB}
\eeqa
This shows that the supersymmetry preserved by the full theory
should be (\ref{BQ}).
We also note that the boundary (\ref{Bb}) is invariant under the
vector R-rotation whereas the axial R-rotation rotates the phase
$\e^{i\beta}$ defining the boundary (\ref{Bb}).

One can develop the boundary superfield formalism identically as in
the case of A-boundary.
B-type supersymmetry transformation
(with parameter $\epsilon_+=-\epsilon_-=\epsilon$,
$\bepsilon=\bepsilon_+=-\bepsilon_-$)
of the superfields is given by
$\delta=\epsilon\cQ-\bepsilon\bcQ$.
A bulk superfield restricted to the boundary (\ref{bosb})-(\ref{Bb})
is a boundary superfield.
The boundary R-rotation comes from the vector R-rotation in the bulk.
A bulk chiral superfield
restricted on the B-boundary is a boundary chiral superfield.

\subsection{$(1,1)$ Superspace and its Boundary}

It is useful also to introduce
the $(1,1)$ superspace and its boundary.
The $(1,1)$ superspace can be defined as a subspace
of the $(2,2)$ superspace:
\beq
\theta^{\pm}=i\theta^{\pm}_1,~~~\mbox{$\theta_1^{\pm}$ real}.
\eeq
(Again, there is a freedom to change the phase $i$ to
$i\e^{i\nu_{\pm}}$ but we set $\nu_{\pm}=0$ for simplicity.)
This subspace can also be defined by the equations
$\theta^{\pm}+\btheta^{\pm}=0$
which are preserved by the differential operators
\beqa
&&{\cal Q}^1_{\pm}:={\cal Q}_{\pm}
+\overline{{\cal Q}}_{\pm}
=-i{\partial\over\partial\theta_1^{\pm}}
+2\theta_1^{\pm}\partial_{\pm},
\\
&&D^1_{\pm}:=D_{\pm}
+\overline{D}_{\pm}
=-i{\partial\over\partial\theta_1^{\pm}}
-2\theta_1^{\pm}\partial_{\pm},
\eeqa
These obey the anti-commutation relations
such as $\{{\cal Q}^1_{\pm},{\cal Q}^1_{\pm}\}=-4i\partial_{\pm}$,
$\{D^1_{\pm},D^1_{\pm}\}=4i\partial_{\pm}$, and
$\{{\cal Q}^1_{\alpha},D^1_{\beta}\}=0$.

The boundary of $(1,1)$ superspace can be defined as
the subspace with
$x^0=0,\pi$ and $\theta_1^+=\pm\theta_1^-$. We take here the plus sign,
$\theta_1^+=\theta_1^-$,
so that both A-boundar and B-boundary (with trivial phases)
of the $(2,2)$ superspace includes this ${\cal N}=1$ boundary as
the subspace
\beq
\theta+\btheta=0.
\label{n1sub}
\eeq
This subspace is preserved by the differential operators
\beqa
&&\cQ^1:=\cQ_+^1+\cQ_-^1=\cQ+\bcQ
=-i{\partial\over\partial\theta_1}+2\theta_1\partial_0,\\
&&D^1:=D_+^1+D_-^1=D+\bD
=-i{\partial\over\partial\theta_1}-2\theta_1\partial_0,
\eeqa
where $\theta_1:={\rm Im}\,\theta$ is the fermionic coordinate of the
${\cal N}=1$ boundary.

It is straighforward to show that for an ${\cal N}=2$ boundary superfield
${\cal F}$ and for a boundary Fermi superfield ${\mit\Psi}$, we have the
identities
\beqa
&&\int\dd\theta\dd\btheta\,{\cal F}=-{i\over 2}\int\dd\theta_1
\left[(D-\bD){\cal F}\right]_1,\\
&&\int\dd\theta\,\,{\mit\Psi}\Bigr|_{\btheta=0}
=-i\int\dd\theta_1[{\mit\Psi}]_1,
\eeqa
where $[-]_1$ stands for the restriction on the ${\cal N}=1$
subspace (\ref{n1sub}).

\section{A Linear Model: An Example}\label{sec:AD}

In this section, we consider the supersymmetric version of
the D1-brane in the complex plane $\C$
which was introduced for the bosonic case in Section~\ref{sec:intro}.
We first take the standard approach to the worlsheet theory based
on supersymmetric boundary condition. There we
identify the chiral parameter of the theory.
We then move on to construct ``a linear model'' of such a
D-brane. The basic idea is to introduce boundary degrees of freedom
and boundary interactions that impose the boundary condition
and also induce the Wilson line term.
We construct the action so that the chiral parameter enters into
a boundary F-term.

The supersymmetric worldsheet theory includes a Dirac fermion fields
$\psi_{\pm}$, $\opsi_{\pm}$ in addition to the complex scalar field
$\phi$.
The action of the system is given by
\beq
S={1\over 2\pi}\int \dd^2x \left(\,
|\partial_{0}\phi|^2-|\partial_1\phi|^2
+{i\over 2}\opsi_-(\lrd_{\!\!\!0}+\lrd_{\!\!\!1})\psi_-
+{i\over 2}\opsi_+(\lrd_{\!\!\!0}-\lrd_{\!\!\!1})\psi_+
\,\right),
\label{cact}
\eeq
where
$\opsi\lrd_{\!\!\!\mu}\psi
:=\opsi\partial_{\mu}\psi-(\partial_{\mu}\opsi)\psi$.
If the worldsheet has no boundary, the action is invariant under
the $(2,2)$ supersymmetry transformations.
This is manifest if we express the action in the $(2,2)$
superspace.
Let $\Phi$ be the chiral superfield which has an expansion
\beq
\Phi=\phi(y)+\theta^{\alpha}\psi_{\alpha}(y)
+\theta^+\theta^-F(y).
\eeq
The action (\ref{cact}) is obtained from
\beq
S={1\over 2\pi}\int \dd^2x\,\dd^4\theta \,\overline{\Phi}\Phi,
\eeq
after an appropriate partial integration
and elimination of the auxiliary field $F$ by its equation of motion.

\subsection{The Boundary Condition}

We now consider the D-brane located at the circle $|\phi|^2=c$
in this supersymmetric theory.
We thus formulate the theory on the left half plane
$\Sigma=\R\times (-\infty,0]$,
and we will find a boundary condition at $\partial \Sigma$
so that the theory
is invariant under A-type supersymmetry.
In the bosonic theory,
the D-brane was represented by 
the boundary condition (\ref{bcon}) at $x^1=0$.
We claim that the boundary condition in the supersymmetric theory is
\beq
~~~~~~\bPhi\Phi=c~~~~\mbox{at A-boundary}.
\label{bcA}
\eeq
By ``at A-boundary'', we mean {\it at A-boundary of the
$(2,2)$ superspace}: $x^1=0$,
$\theta^+=-\btheta^-=\theta$ and $\btheta=\btheta^+=-\theta^-$
(we set $\e^{i\alpha}=1$ in (\ref{Ab})).
In this way of writing, the condition itself is manifestly
invariant under A-type supersymmetry.

In terms of the component fields, the condition (\ref{bcA})
is expressed as
\beq
\begin{array}{l}
|\phi|^2=c,\\[0.1cm]
i\bphi\lrd_{\!\!\!1}\phi+\opsi_+\psi_+-\opsi_-\psi_-
+\overline{F}\phi+\bphi F=0,\\[0.1cm]
\bphi\psi_-+\opsi_+\phi=0,\\[0.1cm]
\opsi_-\phi+\bphi\psi_+=0,
\end{array}
~~~~~\mbox{at $x^1=0$.}
\label{bcAcomp}
\eeq
We note that the first equation is identical to the Dirichlet
boundary condition for $|\phi|$
in (\ref{bcon}) while the second equation generalizes the
Neumann boundary condition for $\arg(\phi)$ in (\ref{bcon}).
It is straightforward to show that the action
(\ref{cact}) plus the auxiliary term ${1\over 2\pi}\int \dd^2 x |F|^2$
is invariant under A-type supersymmetry
$\delta=\epsilon\bcQ-\bepsilon\cQ$ which acts on the component fields as
\beq
\begin{array}{l}
\delta\phi=\epsilon\psi_--\bepsilon\psi_+,\\
\delta\psi_+=\epsilon(2i\partial_+\phi+F),\\
\delta\psi_-=\bepsilon(-2i\partial_-\phi+F),\\
\delta F=-2i\epsilon\partial_+\psi_-
-2i\bepsilon\partial_-\psi_+.
\end{array}
~~~
\begin{array}{l}
\delta\bphi=\epsilon\bpsi_+-\bepsilon\bpsi_-,\\
\delta\bpsi_+=\bepsilon(-2i\partial_+\bphi+\overline{F}),\\
\delta\bpsi_-=\epsilon(2i\partial_-\bphi+\overline{F}),\\
\delta \overline{F}=-2i\bepsilon\partial_+\bpsi_-
-2i\epsilon\partial_-\bpsi_+.
\end{array}
\label{susycomp}
\eeq
Also, one can show that the equation of motion
remains the same as the standard one
\beq
(\partial_0^2-\partial_1^2)\phi=0,
~~(\partial_0\pm \partial_1)\psi_{\mp}=0,
~~F=0,
\label{eoM}
\eeq
under the boundary condition (\ref{bcAcomp}).\footnote{We
notice a slight discrepancy of (\ref{bcAcomp}) from the condition
given in \cite{HIV} which would require
$\bphi\partial_1\phi-\partial_1\bphi\phi=0$.
This is because the requirements (3.2) and (3.3) in \cite{HIV}
was too strong. The most general condition for the variation is
an average of (3.2) and (3.3) but not the separate ones.}

\subsection{The Boundary Term}

Due to the boundary condition $|\phi|^2=c$ with non-zero $c$,
\beq
\varphi:=\arg(\phi)~~~\mbox{at $x^1=0$}
\eeq
is well-defined up to $2\pi$ shifts. 
Then, it is possible to add to the action (\ref{cact})
the following boundary term
\beq
S_{a}
=\int\limits_{\partial\Sigma}\,
{a\over 2\pi}\partial_0\varphi\,\dd x^0.
\label{aterm}
\eeq
Since it is a total derivative in the boundary coordinate, it is a
topological term. In particular
it cannot break the supersymmetry of the system.
Thus the system with the action $S+S_{a}$ is still
invariant under A-type supersymmetry.
The equation of motion also remains the same as (\ref{eoM}).

This boundary term represents the interaction of the
open string end points and the $U(1)$ gauge field on the D-brane
which has holonomy $\e^{ia}$ along the worldvolume
$S^1$.

\subsection{Renormalization and R-Anomaly}\label{subsec:ra}

\subsubsection*{\it Renormalization of $c$}

As mentioned in the introduction,
the boundary condition breaks the scale invariance
of the bulk theory and the constant $c$ runs as the scale is varied.
The renormalization group flow for the D-brane location
was found in \cite{Leigh}
to be the mean curvature flow:
Let us consider a D-brane
whose worldvolume is embedded in the space-time
by the map $f^I(\zeta^{\alpha})$ (where $I$ and $\alpha$ 
are the space-time and the worldvolume indices).
The one-loop beta functional
for the embedding function $f^I(\zeta^{\alpha})$
is given by
\beq
\beta^I=\mu{\dd\over \dd\mu}f^I=-h^{\alpha\beta}K^I_{\alpha\beta},
\eeq
where $h_{\alpha\beta}$
is the induced metric and $K^I_{\alpha\beta}$
is the extrinsic curvature that appears in the normal
coordinate expansion
$f^I=\partial_{\alpha}f^I\zeta^{\alpha}+{1\over 2}K^I_{\alpha\beta}
\zeta^{\alpha}\zeta^{\beta}+\cdots$.
In the present case, the embedding function for our circle $|\phi|^2=c$
is given by
$x=\sqrt{2c}\cos\theta$, $y=\sqrt{2c}\sin\theta$ where
$x$ and $y$ are the normal coordinates on the complex $\phi$-plane,
$\phi=(x+iy)/\sqrt{2}$, and $\theta$ is the angular coordinates of
the circle.
This shows that
$h^{\theta\theta}=1/2c$ and, say at $\theta=0$,
$K^x_{\theta\theta}=-\sqrt{2c}$,
$K^y_{\theta\theta}=0$.
Thus, we have
${\mu}{\dd\over \dd\mu}\sqrt{2c}=1/\sqrt{2c}$.
In other words, the parameter $c$ at the cut-off scale
$\Lambda_{\rm UV}$ and the one at a lower energy scale $\mu$
are related by
\beq
c(\Lambda_{\rm UV})=c(\mu)+\log(\Lambda_{\rm UV}/\mu),
\label{run}
\eeq
at the one-loop level.
This is the story for the bosonic model, but
the fermions does not affect the running of $c$ at the one-loop level,
as in the case of the RG flow of the metric in the bulk
non-linear sigma models \cite{AFM}.

\subsubsection*{\it Axial Anomaly}

There is a related quntum effect; the anomaly of the boundary
R-symmetry.
The bulk theory is invariant under both vector and axial R-rotations.
As mentioned in the previous section, A-boundary
is broken by the vector R-rotation but is preserved by
the axial R-rotation.
The axial rotation (with the trivial R-charge for $\Phi$)
acts trivially on the bosonic fields $\phi$ and $F$
but non-trivially on the fermions as
\beq
\psi_{\pm}\to\e^{\mp i\gamma}\psi_{\pm},~~~
\opsi_{\pm}\to\e^{\pm i\gamma}\opsi_{\pm},
\label{axi}
\eeq
and it indeed preserves the boundary condition (\ref{bcAcomp}).
Thus, the boundary R-rotation that comes from the axial R-rotation
is a symmetry of the classical theory. However, in the quantum theory
it is broken by an anomaly. This can be seen by counting,
as follwos, the number of
fermion zero modes in a topologically non-trivial backgroun.

\newcommand{\bw}{\overline{w}}

We note from (\ref{axi}) that $\psi_-$ and $\opsi_+$ has R-charge $1$
while $\psi_+$ and $\opsi_-$ has R-charge $-1$.
Thus, we are interested in the index which is the difference of the
number of $(\psi_-,\opsi_+)$-zero modes and that of
$(\psi_+,\opsi_-)$-zero modes.

Let $\Sigma$ be the Euclidean left half plane
${\rm Re}(z)\leq 0$ with the canonically flat metric
$\dd s^2=|\dd z|^2$.
We would like to count the above index
in a background $\phi$ in which the worldsheet
$\Sigma$ is mapped to the complex
plane so that the image of the boundary $\partial\Sigma$
(the imaginary axis ${\rm Re}(z)=0$) winds $k$-times around
the circle $|\phi|^2=c$ where the D-brane is located.
The left half-plane is mapped
by the conformal map $w=(1+z)/(1-z)$ to the unit disk $|w|\leq 1$
where the boundary ${\rm Re}(z)=0$ is mapped to
the disk boundary $|w|=1$ with the infinity mapped to $w=-1$.
Now, one can choose
the configuration to be
\beq
\phi(z,\bz)=\sqrt{c}w^k.
\label{config}
\eeq
We note that the positive and the negative chirality spinors are identified
along the boundary $\partial\Sigma$ as
$(\dd z)^{1\over 2}=(\dd \bz)^{1\over 2}$.
\footnote{The reason we do not start from ``$\Sigma$ $=$ the flat disk''
is that the boundary would have an extrinsic curvature
and the Wick rotation of the boundary condition on the fermions
would not be straightforward, the point emphasized in \cite{HIV}.
We could of course have started from the disk with the
hemi-sphereical metric. It is easy to see that the result is the same
as the one given below.} 
In terms of the coordinate $w$ this is translated as
$(\dd w)^{1\over 2}/(w+1)=(\dd\bw)^{1\over 2}/(\bw+1)$ along $|w|=1$.
The boundary conditions on the fermions
\beq
\bphi\psi_-+\opsi_+\phi=0,~~~
\opsi_-\phi+\bphi\psi_+=0,
\label{bcf}
\eeq
should be understood under such an identification.
The zero modes obey the Cauchy-Riemann equations ---
$\psi_-$ and $\opsi_-$ are holomorphic in $z$ or $w$ and
$\psi_+$ and $\opsi_+$ are anti-holomorphic ---
and they must be regular in the disc $|w|\leq 1$.
Thus, they can be expanded as
\beqa
&&\psi_-=\sum_{n=0}^{\infty}c_nw^{n}
\left(\dd w\right)^{1\over 2},
~~~
\opsi_-=\sum_{n=0}^{\infty}b_nw^{n}
\left(\dd w\right)^{1\over 2},\\
&&\psi_+=\sum_{n=0}^{\infty}\overline{c}_n\bw^{n}
\left(\dd \bw\right)^{1\over 2},
~~~
\opsi_+=\sum_{n=0}^{\infty}\overline{b}_n\bw^{n}
\left(\dd \bw\right)^{1\over 2}.
\eeqa
The boundary condition (\ref{bcf}) then requires
\beqa
&&(1+\e^{i\sigma})\sum_{n=0}^{\infty}
c_n\e^{i(n-k)\sigma}
+(1+\e^{-i\sigma})\sum_{n=0}^{\infty}
\overline{b}_n\e^{-i(n-k)\sigma}=0,\\
&&(1+\e^{i\sigma})\sum_{n=0}^{\infty}
b_n\e^{i(n+k)\sigma}
+(1+\e^{-i\sigma})\sum_{n=0}^{\infty}
\overline{c}_n\e^{-i(n+k)\sigma}=0,
\eeqa
where $\sigma$ is the angular part of the polar coordinates
$w=|w|\e^{i\sigma}$ so that $\phi=\sqrt{c}\e^{ik\sigma}$ along the
boundary.
It is easy to see that there is only a trivial solution
$b_n=\overline{c}_n=0$ for the second equation
but there are $2k$ non-singular solutions to
the first one; $c_{i-1}+\overline{b}_{2k-i}=0$
($i=1,2,\ldots,2k$).
Thus the index we wanted to know is $2k$ in the background (\ref{config}).

This means that the path-integral measure changes as
\beq
{\cal D}\psi_{\pm}{\cal D}\opsi_{\pm}\longrightarrow
\e^{-2ki\gamma}
{\cal D}\psi_{\pm}{\cal D}\opsi_{\pm}
\eeq
under the R-rotation (\ref{axi}).
Namely, the classical R-symmetry is anomalously broken.
We note here that the boundary term $S_{a}$
yields the following path-integral weight in this background;
\beq
\exp\left(i\int_{\partial\Sigma}{a\over2\pi}\dd\varphi\right)
=\exp(ika).
\eeq
Thus, the effect of the R-rotation is the shift of $a$
as
\beq
a\longrightarrow a-2\gamma.
\label{shift}
\eeq
This also shows that the parameter $a$ is not actually a physical parameter
of the present theory but can be absorbed
by a field redefinition.

\subsubsection*{\it The Chiral Parameter}

The one-loop running (\ref{run}) of the parameter $c$ and the R-anomaly
(\ref{shift}) suggests that the parameters $c$ and $a$ are superpartners
of each other and can be combined into a complex parameter.
The precise combination can be found by identifying
the instantons and computing the action.
A configuration such that $\phi$ is holomorphic and $F=0$
preserves half of the supersymmetry. (See the
$\epsilon$-variation in the Euclidean version of (\ref{susycomp}).)
The configuration (\ref{config}) is an example of such an instanton.
The Euclidean action in such a background is
\beqa
S_E&=&{1\over 2\pi}\int\limits_{\Sigma}
\left(2|\partial_z\phi|^2+2|\partial_{\bz}\phi|^2\right)\dd^2z
-i{a\over 2\pi}\int\limits_{\partial\Sigma}\dd\varphi
\nn\\
&=&{c\over 2\pi}\int\limits_{|w|\leq 1}
2|\partial_w w^k|^2 \dd^2w-i{a\over2\pi}2\pi k
=(c-ia)k.
\eeqa
This shows that the right complex combination is
\beq
s=c-ia.
\label{s}
\eeq

\subsection{A Linear Model}

We now construct a linear model
of this A-type D-brane.
The general structure of the worldsheet
action we would like to have is as follows.
It is a sum of two parts,
\beq
S_{\it tot}=S(\Phi)+S_{\it bd}(U,\Phi)
\eeq
where $S(\Phi)$ involves only the bulk field $\Phi$
and includes the bulk action (\ref{cact}) while
$S_{\it bd}(U,\Phi)$ is the boundary interaction of $\Phi$
and a boundary superfield $U$
that imposes the boundary condition (\ref{bcA}) at low enough energies.
The boundary condition on $\Phi$ is not
something we impose by hand at the beginning,
but is regarded as derived through the interaction with the
boundary fields.
We in particular require $S_{\it tot}$
to be supersymmetric without using boundary condition on $\Phi$
nor its equation of motion.
Also, we would like
$S_{\it bd}(U,\Phi)$ to be manifestly
supersymmetric so that the parameter $s$ in (\ref{s}) 
appears in the boundary superpotential.
This requires $S(\Phi)$ to be supersymmetric by itself.

We start with finding the boundary term.
Let us introduce a real bosonic boundary superfield
\beq
U=u+\theta\bchi-\btheta\chi+\theta\btheta E.
\eeq
The Lagrangian
\beq
L_{\it bd}^{(1)}=\int \dd\theta\dd\btheta\,
(\bPhi\Phi-c)U,
\eeq
imposes the boundary condition $\bPhi\Phi=c$ via
the equation of motion for $U$.
This however does not lead to the $a$-term (\ref{aterm})
nor does it make clear that $s=c-ia$ is a chiral parameter.
We note that one can replace
the term $\int\dd\theta\dd\btheta\,(-c)U$
by ${1\over 2}\int\dd\theta c\bD U+c.c.$.
This motivates us to define the ``fieldstrength'' $\Upsilon$ of $U$
as
\beq
\Upsilon:=\bD \,U
=\chi+\theta(E+i\partial_0 u)-i\theta\btheta\partial_0\chi,
\eeq
which is a boundary Fermi superfield, $\bD\Upsilon=0$.
Now we try the following Lagrangian
\beq
L_{\it bd}^{(2)}=\int\dd\theta\dd\btheta\,
\bPhi\Phi U
+{\rm Re}\int\dd\theta\, s\Upsilon.
\eeq
in which $s$ is manifestly a chiral parameter.
In terms of the component fields the boundary superpotential term
can be written as
\beq
{\rm Re}\int\dd\theta\, s\Upsilon
=c\,E\,+\,a\,\partial_0u.
\eeq
We indeed find an $a$-term. However, we should note that
$u$ has no relationship with $\varphi=\arg(\phi)$
at this stage.
Furthermore, if $u$ is a single valued field
(an ordinary field with values in $\R$),
the term $a \partial_0 u$ can be set equal to zero
since it is a total derivative.
We may declare that $u$ has the same periodicity as $\varphi$
or that $u-\varphi$ is a single valued field.
However, that would make
the term $\int \dd\theta\dd\btheta\,\bPhi\Phi U$ ill-defined;
it changes its value as $u\to u+2\pi$.

Putting aside this problem, let us try to determine the
term $S(\Phi)$.
We would like it to be an action with the bulk part
(\ref{cact})
that is invariant under A-type supersymmetry without using
a boundary condition.
It is straighforward to see that the following meets such a requirement;
\beqa
S_A(\Phi)&=&{1\over 2\pi}\int\limits_{\Sigma}
\left(\,
|\partial_{0}\phi|^2-|\partial_1\phi|^2
+{i\over 2}\opsi_-(\lrd_{\!\!\!0}+\lrd_{\!\!\!1})\psi_-
+{i\over 2}\opsi_+(\lrd_{\!\!\!0}-\lrd_{\!\!\!1})\psi_+
+|F|^2
\,\right)\dd^2x
\nn\\
&&+{1\over 4\pi}\int\limits_{\partial\Sigma}
\Bigl(\,\partial_1|\phi|^2+i(\overline{F}\phi-\bphi F)\,\Bigr)\dd x^0.
\label{SPhi}
\eeqa
However, the second term (boundary term) is non-vanishing
even if we use the boundary condition $\bPhi\Phi=c$.
This is something we do not want.

Thus,
the candidate action
$S_A(\Phi)+{1\over 2\pi}\int_{\partial\Sigma}L_{\it bd}^{(2)}\dd x^0$
is invariant under A-type supersymmetry but has two problems;
$S_A(\Phi)$ contains unwanted boundary interactions
and the term $\int_{\partial\Sigma}L_{\it bd}^{(2)}\dd x^0$
either is ill-defined or lacks the $a$-term.
Fortunately, both of these problems can be cancelled by
addition of the following boundary term in the Lagrangian
\beq
{\mit\Delta}L_{bd}
=-\int\dd\theta\dd\btheta\,\bPhi\Phi{\rm Im}\log\Phi
={i\over 2}\int \dd\theta\dd\btheta\,\bPhi\Phi(\log\Phi-\log\bPhi).
\label{save}
\eeq
The unwanted boundary interaction in $S_A(\Phi)$ is precisely cancelled
by this term. Also, $L_{\it bd}^{(2)}+{\mit\Delta}L_{\it bd}$
contains the term
$\int\dd\theta\dd\btheta\,(u-\varphi) \bPhi\Phi$,
which requires $u-\varphi$ to be a single valued field.
Then, the $a$-term for $u$ becomes the $a$-term for $\varphi$;
\beq
{a\over 2\pi}\int\limits_{\partial\Sigma}\partial_0u\dd x^0
={a\over 2\pi}\int\limits_{\partial\Sigma}\Bigl\{\partial_0\varphi
+\partial_0(u-\varphi)\Bigr\}\dd x^0
={a\over 2\pi}\int\limits_{\partial\Sigma}\partial_0\varphi\dd x^0.
\label{reda}
\eeq
It may be appropriate to explain on the single-valuedness
of the boundary interaction $L^{(2)}_{\it bd}+{\mit\Delta}L_{\it bd}$,
in particular the term $\bPhi\Phi(U-{\rm Im}\log \Phi)$.
We can express the field $\Phi$ as $\Phi=\e^{\Psi}$ and consider
$\Phi$ as a gauge invariant field where the gauge symmetry is $\Z$
which acts on $\Psi$ as $\Psi\to \Psi+2\pi in$.
We then consider $U$ as the gauge field on which $\Z$ acts by
$U\to U+2\pi n$. Then, $u-\varphi$ is gauge invariant and
$\partial_0u\dd x^0$ can really be considered as the fieldstrength
of this gauge symmetry.

To summarize, the total action we were looking for is
\beqa
S_{\it tot}&=&S_A(\Phi)
+{1\over 2\pi}\int\limits_{\partial\Sigma}
\dd x^0
\left[~\int\dd\theta\dd\btheta\,
\bPhi\Phi(U-{\rm Im}\log\Phi)
+{\rm Re}
\int \dd\theta
\,s\Upsilon~\right],
\nn\\
\label{Action}
\eeqa
where $S_A(\Phi)$ is given by (\ref{SPhi}).
The boundary term appears highly non-linear in $\Phi$
but the essential non-linearity resides only in $\bPhi\Phi\varphi$
which is absorbed by a redefinition of
$u$ that simply yields the $a\partial_0\varphi$ term (\ref{aterm})
when $|\phi|\ne 0$.

We note that the above boundary interaction reproduces the
correct one-loop running (\ref{run}) of the parameter $c$.
Up to the topological term, the bosonic part of the action
(which is the relevant part in this discussion)
is given by
\beq
S={1\over 2\pi}\int\limits_{\Sigma} \dd^2x 
\left(\,
|\partial_{0}\phi|^2-|\partial_1\phi|^2\,\right)
+{1\over 2\pi}\int\limits_{\partial\Sigma}
\dd x^0\left\{E(|\phi|^2-c)+iu'\,\bphi\lrd_{\!\!\!1}\phi\right\}
\eeq
where $u'=u-\varphi$ is the single valued field.
In the effective action at the energy scale $\mu\ll \Lambda_{\rm UV}$,
$|\phi|^2$ in $E(|\phi|^2-c)$ is shifted by
$\langle |\phi|^2\rangle_{\mu}$,
the one-loop momentum integral
in the range $\mu\leq |k|\leq\Lambda_{\rm UV}$.
In the present case one real component of $\phi$ obeys the Neumann boundary
condition and the other obeys Dirichlet. Thus, the
one point function has the same divergence as the ordinary one
point function in the bulk.
Thus, the shift is by
\beq
\int\limits_{\mu\leq |k|\leq \Lambda_{\rm UV}}
{\dd^2k\over (2\pi)^2}{2\pi\over k^2}=\log(\Lambda_{\rm UV}/\mu).
\label{onel}
\eeq
This divergence (\ref{onel})
is absorbed exactly by giving the scale dependence
of $c$ as in (\ref{run}).

We stress that the most important aspect of this
formulation is that
the parameter $s=c-ia$ enters into the boundary F-term.
Any correction to the boundary F-term has to be holomorphic in
$s$, periodic under $2\pi i$ shifts of $s$, and of boundary R-charge $1$
if we assign R-charge $2$ to $\e^{-s}$.
This excludes any perturbative renormalization except at one loop.
Also, if we require the correction to be small at large $s$,
we see that no correction is possible at all.
In more general examples we will consider in the next section,
we will always see this perturbative non-renormalization
theorem (except at one-loop).
However, non-perturbative non-renormalization still holds in some cases
but fails in some other cases.

\section{A-Type D-Branes in Linear Sigma Model}\label{sec:LinA}

In this section, we apply the construction of the linear model
to gauge theories.
This enables us to define A-type D-branes in toric sigma
models so that the D-brane location and the Wilson lines,
combined into complex chiral parameters, enter into
a boundary F-term.
We will also find the dual description of the D-brane
in the mirror Landau-Ginzburg model.

\subsection{Supersymmetric Gauge theory}\label{subsec:gauge}

Here we record the
basic facts on supersymmetric gauge theory in the bulk.
We consider the simplest example: $U(1)$ gauge theory
with a single chiral matter field of charge $1$.
The gauge transformation of the vector superfield $V$ and the chiral matter
field $\Phi$ is given by
\beq
V\longrightarrow V-iA+i\overline{A},
~~~
\Phi\longrightarrow\e^{iA}\Phi,
\label{GaugeT}
\eeq
where $A$ is a chiral superfield.
We usually partially fix the gauge 
so that the vector superfield takes the form
\beqa
V&=&\theta^-\btheta^-(v_0-v_1)+\theta^+\btheta^+(v_0+v_1)
-\theta^-\btheta^+\sigma-\theta^+\btheta^-\bsigma
\nn\\
&&
+i\theta^-\theta^+(\btheta^-\blambda_-+\btheta^+\blambda_+)
+i\btheta^+\btheta^-(\theta^-\lambda_-+\theta^+\lambda_+)
+\theta^-\theta^+\btheta^+\btheta^- D.
\label{WZgauge}
\eeqa
$v_0$ and $v_1$ define a one-form field,
$\sigma$ is a complex scalar field,
$\lambda_{\pm}$ and $\blambda_{\pm}$ define
a Dirac fermion, and $D$ is a real scalar field.
This is called the Wess-Zumino gauge and
the residual gauge symmetry
is the one with $A=\alpha(x)$ which rotates the phase of $\Phi$ and
transforms $v_{\mu}$ as
\beq
v_{\mu}(x)\to v_{\mu}(x)-\partial_{\mu}\alpha(x).
\eeq
The supersymmetry variation
$\delta=\epsilon_+{\cal Q}_--\epsilon_-{\cal Q}_+
-\bepsilon_+\overline{\cal Q}_-+\bepsilon_-\overline{\cal Q}_+$
does not in general preserve
the Wess-Zumino gauge.
In order to find the
supersymmetry transformation of the component fields
$\sigma$, $\lambda_{\pm}$, $v_{\mu}$ and $D$,
we need to amend it with a gauge transformation that brings
$\delta V$ back into the Wess-Zumino gauge.
It turns out that
the required gauge transformation is the one with
\beq
A=i\theta^+\left(\bepsilon_+\bsigma+\bepsilon_-(v_0+v_1)\right)
-i\theta^-\left(\bepsilon_-\sigma+\bepsilon_+(v_0-v_1)\right)
+\theta^+\theta^-\left(\bepsilon_-\blambda_+-\bepsilon_+\blambda_-\right)
+\cdots,
\eeq
where $+\cdots$ are the derivative terms to make $A$ chiral.
In this way, we find that
the supersymmetry transformation of the component fields of $V$ is
\bea
&&\delta(v_0\pm v_1)=i\oep_{\pm}\lambda_{\pm}
+i\eps_{\pm}\overline{\lambda}_{\pm},
\nn\\
&&\delta \sigma =-i\oep_{+}\lambda_{-}-i\eps_{-}\overline{\lambda}_{+},
\nn\\
&&\delta D =-\oep_{+}\partial_{-}\lambda_{+}-\oep_{-}\partial_{+}
\lambda_{-}+\eps_{+}\partial_{-}\overline{\lambda}_{+}+\eps_{-}\partial_{+}
\overline{\lambda}_{-},
\nn\\
&&\delta \lambda_{+}= i \eps_{+}(D+i v_{01})
+2\eps_{-}\partial_{+}\overline{\sigma},
\nn\\
&&\delta \lambda_{-}=i \eps_{-}(D-i v_{01})
+2\eps_{+}\partial_{-}\sigma,
\nn
\eea
while that of $\Phi$ is
\beqa
&&\delta\phi=\epsilon_+\psi_--\epsilon_-\psi_+,
\nn\\
&&\delta\psi_+=i\bepsilon_-(D_0+D_1)\phi+\epsilon_+F-\bepsilon_+\bsigma\phi,
\nn\\
&&\delta\psi_-=-i\bepsilon_+(D_0-D_1)\phi+\epsilon_-F
+\bepsilon_-\sigma\phi,
\nn\\
&&\delta F=-i\bepsilon_+(D_0-D_1)\psi_+-i\bepsilon_-(D_0+D_1)\psi_-
\nn\\
&&~~~~~~~+\bepsilon_+\bsigma\psi_-+\bepsilon_-\sigma\psi_+
+i(\bepsilon_-\blambda_+-\bepsilon_+\blambda_-)\phi.~~~~~
\nn
\eeqa
The superfield
\beq
\Sigma:=\overline{D}_+D_-V
\eeq
is invariant under the gauge transformation $V\to V+i(\overline{A}-A)$.
It is a twisted chiral superfield
which is expressed in the Wess-Zumino gauge as
\beq
\Sigma=\sigma(\widetilde{y})+i\theta^+\blambda_+(\widetilde{y})
-i\btheta^-\lambda_-(\widetilde{y})
+\theta^+\btheta^-(D-iv_{01})(\widetilde{y}).
\eeq
where $v_{01}$ is the field-strength
$v_{01}:=\partial_0v_1-\partial_1v_0$.
The superfield $\Sigma$ is called the {\it super-field-strength}
of $V$.

We consider
the following gauge invariant action
\beq
S={1\over 2\pi}\int\limits_{\Sigma} \left[\int \dd^4\theta\,
\left(\overline{\Phi}\e^V\Phi
-{1\over 2e^2}\overline{\Sigma}\Sigma\right)
+{\rm Re}\int d^2\widetilde{\theta}\,(-t\Sigma)~
\right]\dd^2x,
\label{gaction}
\eeq
where
\beq
t=r-i\theta
\eeq
is a dimensionless twisted chiral parameter;
$r$ is the Fayet-Illiopoulos parameter
and $\theta$ is the Theta angle.
This Lagrangian is invariant
under $(2,2)$ supersymmetry when the worldsheet $\Sigma$ has no
boundary.
In terms of the component fields
this is written as
\beqa
S\!\!&=&\!\!{1\over 2\pi}\int\limits_{\Sigma}\Biggl[\,
-D^{\mu}\bphi D_{\mu}\phi
+{i\over 2}\opsi_{-}(\lrD_{\!\!0}+\lrD_{\!\!1})\psi_{-}
+{i\over 2}\opsi_{+}(\lrD_{\!\!0}-\lrD_{\!\!1})\psi_{+}+D|\phi|^{2}+|F|^2
\nn\\[-0.15cm]
&&~~~~~~~-|\sigma|^{2}|\phi|^{2}-\opsi_{-}\sigma\psi_{+}
-\opsi_{+}\bsigma\psi_{-}-i\bphi\lambda_{-}\psi_{+} 
+i\bphi\lambda_{+}\psi_{-}
+i\opsi_{+}\blambda_{-}\phi -i\opsi_{-}\blambda_+\phi
\nn\\[0.15cm]
&&~~~~~~~+\frac{1}{2e^{2}}\left(
-\partial^{\mu}\bsigma\partial_{\mu}\sigma
+{i\over 2}\blambda_{-}(\lrd_{\!\!\!0}+\lrd_{\!\!\!1})\lambda_{-}
+{i\over 2}\blambda_{+}(\lrd_{\!\!\!0}-\lrd_{\!\!\!1})\lambda_{+}
+v_{01}^{2}+D^{2}\right)
\nn\\[0.05cm]
&&~~~~~~~-rD+\theta v_{01}~\Biggr]\,\dd^2x,
\label{gact}
\eeqa
where an appropriate partial integration is made.

This theory is super-renormalizable with respect to the dimensionful gauge
coupling constant $e$. The FI parameter $r$ is renormalized
in such a way as
\beq
r(\Lambda_{\rm UV})=r(\mu)+\log(\Lambda_{\rm UV}/\mu).
\eeq
The vector R-symmetry is unbroken but the axial R-symmetry is anomalous;
The axial R-rotation shifts the Theta angle as
$\theta\to\theta-2\alpha$.

\subsection{The Boundary Interaction}

Let us now choose the worldsheet $\Sigma$ to be the strip
$\R\times [0,\pi]$.
The $(2,2)$ supersymmetry variantion of the
action (\ref{gact}) is given by a boundary term, as analyzed in
\cite{HIV}.
For A-type supersymmetry one can show that
the combination
\beq
S_A=S+{1\over 2\pi}\int\limits_{\partial\Sigma}
\dd x^0\left[\,{1\over 2}\partial_1|\phi|^2
+{i\over 2}(\overline{F}\phi-\bphi F)
+{i\over 4e^2}(\lambda_-\lambda_+-\blambda_+\blambda_-)
+\theta v_0\,\right]
\label{Sp}
\eeq
has the following simple transformation property;
\beq
\delta S_A={r\over 4\pi}\int\limits_{\partial\Sigma}
\dd x^0\left\{\epsilon(\blambda_++\lambda_-)
-\bepsilon(\blambda_-+\lambda_+)\right\}.
\label{delSp}
\eeq

We would like to construct a linear model for
D-branes in this gauge theory.
As in the case without the gauge field,
we introduce the superfield $U$ on the A-boundary
with the boundary interaction 
as in (\ref{Action}):
\beq
S_{\it boundary}={1\over 2\pi}\int\limits_{\partial\Sigma}
\dd x^0
\left[~\int\dd\theta\dd\btheta\,
\bPhi\e^V\Phi(U-{\rm Im}\log\Phi)
+{\rm Re}
\int \dd\theta
\,s\Upsilon~\right],
\eeq
where $\Upsilon=\bD U$ is the fieldstrength.
For gauge invariance of the first term,
we would like the field $U$ to transform in the same way
as ${\rm Im}\log\Phi$. Thus, it transforms as
\beq
U\longrightarrow U+{1\over 2}(A+\overline{A}),
\label{GaU}
\eeq
under the gauge transformation (\ref{GaugeT}).
In particular, the supersymmetry transformation of the
component field is modified in the Wess-Zumino gauge as
\beqa
&&\delta u=\epsilon\chi-\bepsilon\bchi,
\nn\\
&&\delta \chi=-\bepsilon(E+i(\partial_0 u+v_0))
-i\epsilon\sigma,
\nn\\
&&\delta \bchi=-\epsilon(E-i(\partial_0 u+v_0))
+i\bepsilon\bsigma,\nn\\
&&\delta E=i\epsilon\partial_0\chi+i\bepsilon\partial_0\bchi
-{1\over 2}\epsilon(\lambda_-+\blambda_+)
+{1\over 2}\bepsilon(\blambda_-+\lambda_+).
\nn\\
\eeqa
Under this modified supersymmetry transformation,
the boundary superpotential term
\beq
{1\over 2\pi}\int\limits_{\partial\Sigma}
\dd x^0~{\rm Re}\int\dd\theta
\,s\Upsilon\,=
{1\over 2\pi}\int\limits_{\partial\Sigma}
\dd x^0\left(cE+a\partial_0 u\right)
\label{boundt}
\eeq
is not invariant but varies as
\beq
\delta \left[{1\over 2\pi}\int\limits_{\partial\Sigma}
\dd x^0\,{\rm Re}\int\dd\theta
\,s\Upsilon~
\right]
=-{c\over 4\pi}
\int\limits_{\partial\Sigma}
\dd x^0\left\{\epsilon(\blambda_++\lambda_-)
-\bepsilon(\blambda_-+\lambda_+)\right\}.
\label{nontV}
\eeq
We note that this is proportional to $\delta S_A$ given in
(\ref{delSp}).
Thus, $S_A+S_{\it boundary}$ is invariant under A-type supersymmetry
if and only if
\beq
c=r.
\eeq
The non-trivial variation (\ref{nontV}) comes from the fact that
the Fermi-superfield $\Upsilon$ is not invariant under
the gauge transformation (\ref{GaU}).
This reminds us of another constraint;
The action must be gauge invariant, or it must be invariant
under the residual gauge transformation
$A=\alpha(x)$ in the Wess-Zumino gauge we are in.
The gauge transformation shifts
the component $u$ as $u\to u+\alpha$ while $\chi$ and $E$ are invariant.
Thus the term (\ref{boundt}) is shifted
by ${a\over 2\pi}\int \dd x^0 \partial_0\alpha$.
We also notice that the action $S_A$ is not gauge invariant but
is shifted by
$-{\theta\over 2\pi}\int\dd x^0 \partial_0\alpha$
since $v_{\mu}$ is transformed as $v_{\mu}\to v_{\mu}-\partial_{\mu}\alpha$.
Thus, the action $S_A+S_{\it boundary}$ is gauge invariant
if and only if
\beq
~~~~~a=\theta~~~({\rm mod}~2\pi\Z).
\eeq
Thus, the action
\beq
S_{\it tot}=S_A+S_{\it boundary}
\eeq
is supersymmetric and gauge invariant if and only if
\beq
~~~~~s=t ~~~({\rm mod}~2\pi i\Z).
\eeq
We note that the real part of the condition, $r=c$,
makes the derived boundary condition $|\phi|^2=c$
to be compatible with the D-term constraint $|\phi|^2=r$.
We also note again that no boundary condition on the bulk fields are required
for the supersymmetry.

\subsection{Construction in Linear Sigma Model}

Let us consider a $U(1)$ gauge theory with
$N$ chiral multiplets $\Phi_i$ of charge $Q_i$
where the action is given by
\beq
S={1\over 2\pi}\int\limits_{\Sigma} \left[~\int \dd^4\theta\,
\left(\sum_{i=1}^N\overline{\Phi}_i\e^{Q_iV}\Phi_i
-{1\over 2e^2}\overline{\Sigma}\Sigma\right)
+{\rm Re}\int d^2\widetilde{\theta}\,(-t\Sigma)~
\right]\dd^2x,
\eeq
The theory
reduces at low enough energies to non-linear sigma model
on a certain toric manifold $X$.
For instance, $X=\CP^{N-1}$ if $Q_i=1$ for all $i$;
$X$ is the total space of the line bundle ${\cal O}(-d)$ over $\CP^{N-2}$
if $Q_1=\cdots =Q_{N-1}=1$ and $Q_{N}=-d$;
$X$ is the total space of
${\cal O}(-1)\oplus {\cal O}(-1)$ over $\CP^1$
if $N=4$ and $Q_1=Q_2=1$, $Q_3=Q_4=-1$.
The character of the theory depends on whether the sum of charges
\beq
b_1=\sum_{i=1}^NQ_i
\eeq
is zero or not.
If $b_1\ne 0$, the scale invariance is broken at the one-loop level and the
dimensionless FI parameter $r$ is replaced by a dynamically generated
scale parameter $\Lambda$ by
$r(\mu)=b_1\log(\mu/\Lambda)$.
Also, the axial $U(1)$ R-symmetry is anomalously broken;
the axial R-rotation shifts the Theta angle as
$\theta\to\theta-2b_1\beta$.
The sign of $r$ at the cut-off scale is determined by
$b_1$ and hence the target space is uniquely determined
by the charges $Q_i$.
On the other hand, if $b_1=0$, the scale invariance is preserved at the
one-loop level. In particular the FI-Theta parameter
$t=r-i\theta$ is the dimensionless parameter of the theory.
The toric manifold $X$ in this case is a (non-compact) Calabi-Yau manifold.
We can choose the sign of $r$ as we wish;
both positive and negative $r$ are possible
and the sigma model target space $X$ differs in general.

Let us now formulate the theory on the left half plane
$\Sigma=\R\times [-\infty,0]$.
To fix the expression of the bulk action $S$
in terms of the component field,
we take the obvious
generalization of the standard one (\ref{gact}).
We also define $S_A$ in the same way as in (\ref{Sp}). Then we still have
the simple supersymmetry variation (\ref{delSp}).
Now, we introduce $N$ boundary real superfields $U_i$ 
(with fieldstrength $\Upsilon_i$) and
the boundary interaction term
\beq
S_{\it boundary}={1\over 2\pi}\sum_{i=1}^N\int\limits_{\partial\Sigma}
\dd x^0
\left[~\int\dd\theta\dd\btheta\,
\bPhi_i\e^{Q_iV}\Phi_i(U_i-{\rm Im}\log\Phi_i)
+{\rm Re}
\int \dd\theta
\,s_i\Upsilon_i~\right],
\label{Bount}
\eeq
One can show as before that the total action
$S_{\it tot}=S_A+S_{\it boundary}$ is gauge invariant and supersymmetric
if and only if
\beq
~~~~~\sum_{i=1}^NQ_is_i=t~~~({\rm mod}~2\pi i\Z).
\label{sgcond1}
\eeq
We note that the one-loop renormalization group flows of $s_i$'s and $t$
are compatible with each other:
The former runs as $s_i=\log(\mu/\Lambda)+$constant
while the latter runs as $t=b_1\log(\mu/\Lambda)$
only if $b_1=\sum_{i=1}^NQ_i$ is non-zero.
In any case there are one scale parameter and $N-1$ dimensionless
complex parameters.

It is straightforward to generalize the construction to the case of
higher rank gauge group $U(1)^k=\prod_{a=1}^kU(1)_a$ with $N$ matters
with charge $Q_{ia}$.
The condition for supersymmetry and gauge invariance is
$\sum_{i=1}^NQ_{ia}s_i=t_a$ where $t_a$ is the FI-Theta parameter for
$U(1)_a$.

\subsubsection*{\it Geometric Interpretation}

Let us find out what the above boundary interaction
corresponds to in the non-linear sigma model limit $\mu\ll e$.
Integrating out the boundary superfields $U_i$
yields the boundary condition
\beq
\bPhi_i\e^{Q_iV}\Phi_i=c_i~~~\mbox{at A-boundary}.
\label{Phici}
\eeq
We note that at high enough enrgies
$\mu\gg\Lambda$, $c_i$ are all positive and one can solve the constraint
(\ref{Phici}).
This boundary condition corresponds to a D-brane wrapped on the
$(N-1)$-dimensional torus $T$ in $X$ located at
\beq
|\phi_i|^2=c_i.
\label{constraa}
\eeq
It is easy to see that $T$ is a Lagrangian submanifold of $X$.
The parameter $a_i$ parametrizes the holonomy
of the flat $U(1)$ gauge field on the D-brane
since the boundary term $S_{\it boundary}$ contains the term
\beq
{1\over 2\pi}\sum_{i=1}^N\int\limits_{\partial\Sigma}
a_i\dd\varphi_i,
\eeq
that is obtained through the process (\ref{reda}).
It may appear that the holonomy in the unphysical gauge orbit direction is
$\sum_{i=1}^NQ_ia_i=\theta$ which is non-vanishing
for a non-zero worldsheet Theta angle.
However, it is not the case;
there is also a boundary term
${\theta\over 2\pi}\int_{\partial\Sigma} \dd x^0 v_0$ (see (\ref{Sp})).
At low enough energies, the worldsheet gauge
field $v_{\mu}$ is frozen at
\beq
v_{\mu}={i\over 2}{\sum_{i=1}^NQ_i
\bphi_i\lrd_{\!\!\!\mu}\phi_i\over \sum_{i=1}^NQ_i^2|\phi_i|^2}
=
-{\sum_{i=1}^NQ_ic_i\partial_{\mu}\varphi_i\over
\sum_{i=1}^NQ_i^2c_i},
\eeq
where in the last step we have used the constraint (\ref{constraa}).
Thus, the total holonomy term is
\beq
{1\over 2\pi}\int\limits_{\partial\Sigma}
\left[\,\sum_{i=1}^N a_i\dd\varphi_i
\,-\,\theta\left(\sum_{i=1}^NQ_ic_i\dd\varphi_i\Biggr/
\sum_{i=1}^NQ_i^2c_i\right)\right].
\eeq
It is easy to see that the holonomy in the gauge orbit direction
indeed vanishes if the condition
$\theta=\sum_{i=1}^NQ_ia_i$ in (\ref{sgcond1}) is satisfied.

We recall that the parameters $c_i$ are running coupling constants,
irrespective of whether $b_1=\sum_{i=1}^NQ_i$ is zero or not.
$c_i$ becomes smaller as the energy is reduced.
If $b_1> 0$, the manifold $X$ itself also becomes smaller
at lower energies.
Thus, in this case, there is a chance that the D-brane
stays in the theory at low energies where the sigma model description
breaks down.
We will indeed find in the mirror description
(which is the better description at low energies)
that the D-brane stays in the theory for special values of $s_i$.
If $b_1=0$, the size of the manifold $X$ does not change.
It is expected that, as in the basic example of Section~\ref{sec:AD},
the D-brane disappears from the theory at extreme
low energies.

As usual, it is easy to exclude perturbative renormalization beyond
one-loop level.
We claim that there is no non-perturbative renormalization either
in the case where $b_1>0$..
This follows from the requirement that the correction is small
at small $\Sigma$, at large $s_i$ for any $i$ and at large $t$.

\subsection{Promoting $s_i$ to Chiral Superfields}\label{subsec:promote}

There is actually an interesting generalization of the above
construction. It is to promote the parameters $s_i$ to
boundary chiral superfields.
The gauge symmetry and the supersymmetry is not spoiled even if
we do so, provided the condition (\ref{sgcond1})
is obeyed.
Thus, we make the replacement
\beq
s_i\longrightarrow S_i(Z_1,\ldots,Z_{\ell}),
\label{sifun}
\eeq
where
$Z_{\alpha}$ are boundary chiral superfields and
$S_i(Z_{\alpha})$ are holomorphic functions
obeying
$\sum_{i=1}^NQ_{i}S_i(Z_{\alpha})=t$.
A simple class of such functions are linear ones.
This is motivated by the recent work \cite{AV}.
Let $m_{i}^{\alpha}$ be such that $\sum_{i=1}^NQ_{i}m_i^{\alpha}=0$.
Then one can take
\beq
S_i=\sum_{\alpha=1}^{\ell}m_i^{\alpha}Z_{\alpha}
+s_i,
\label{lin}
\eeq
where $s_i$ are parameters obeying $\sum_{i=1}^NQ_{i}s_i=t$.

When the charges $Q_{i}$ satisfy the condition
$b_{1}=\sum_{i=1}^NQ_{i}=0$, the bulk theory is scale invariant at the
one-loop level and is expected to flow to a non-trivial SCFT in the
infra-red limit.
In such a case, it is natural to ask under what condition the boundary 
interaction does not break the scale invariance.
We recall that $c_i={\rm Re}(s_i)$ is a running  coupling constant:
\beq
c_i(\Lambda_{\rm UV})=c_i(\mu)+\log(\Lambda_{\rm UV}/\mu).
\label{runci}
\eeq
However, this running can be absorbed by the shift of the fields
$Z_{\alpha}$ by a certain condition on $m_i^{\alpha}$'s.
The condition is that there are numbers $\delta_{\alpha}$ such that
\beq
\sum_{\alpha=1}^{\ell}m_i^{\alpha}\delta_{\alpha}=1.
\label{scal}
\eeq
In such a case, the boundary interaction is scale invariant at
the one-loop level.

\subsubsection*{\it Geometric Interpretation and a Constraint
on the Parameters}

Let us see what this boundary interaction corresponds to
in the non-linear sigma model.
The $Z_{\alpha}$-equation of motion yields the constraint
$\sum_{i=1}^Nm_i^{\alpha}\bD U_i=\sum_{i=1}^Nm_i^{\alpha}DU_i=0$
which means
\beq
\sum_{i=1}^Nm_i^{\alpha}U_i={\rm constant}~~~
\mbox{at A-boundary}.
\eeq
Here ``constant at A-boundary'' means that it does not depend on
$x^0,\theta,\btheta$. Namely, the coefficient of $\theta,\btheta$ and
$\theta\btheta$ vanishes and the leading term is independent of $x^0$.
On the other hand, the equation of motion for $\Phi_i$ identifies
${\rm Im}\log\Phi_i$ with $U_i$.
(In the derivation of this statement
the bulk and the boundary terms in the action $S_A$
play an important role.)
Thus, we obtain the following constarint for $\Phi_i$'s:
\beq
\sum_{i=1}^Nm_i^{\alpha}{\rm Im}\log\Phi_i={\rm constant}~~~
\mbox{at A-boundary}.
\label{condIm}
\eeq
The equation of motion for the remaining $U_i$ yields 
the condition
\beq
\bPhi_i\e^{Q_iV}\Phi_i={\rm Re}\left(\sum_{\alpha=1}^{\ell}m_i^{\alpha}
Z_{\alpha}+s_i\right)
~~~\mbox{at A-boundary}.
\label{condabs}
\eeq
The boundary conditions (\ref{condIm})-(\ref{condabs})
are that for a D-brane located at
\beqa
&&|\phi_i|^2=\sum_{\alpha=1}^{\ell}m_i^{\alpha}\zeta_{\alpha}+c_i,
~~~~i=1,\ldots,N;
\label{aff}
\\
&&\sum_{i=1}^Nm_i^{\alpha}\varphi_i={\rm constant},
~~~\alpha=1,\ldots,\ell,
\label{toru}
\eeqa
where $\zeta_{\alpha}$ are real coordinates that can vary.
It is a $(N-1)$ dimensional subspace $L$ which is a fibration over
the $\ell$-dimensional subspace (\ref{aff}) in the $|\phi_i|^2$-plane
with the $(N-\ell-1)$ dimensional torus (\ref{toru})
as its fibre. It is a Lagrangian submanifold of $X$.
We note that $\ell$ of the $c_i$ parameters can be absorbed
by the redefinition of $\zeta_{\alpha}$.
Also, by the constraints (\ref{toru}), the $a_i$'s related by the shifts
by $m_i^{\alpha}$ are physically equivalent.
There are only $(N-\ell-1)$ physical parameters.

When the charges $Q_{ia}$ satisfy $b_{1,a}=\sum_{i=1}^NQ_{ia}=0$,
the target space is a non-compact Calabi-Yau manifold
and the sigma model is scale invariant at the one-loop level
(cooresponding, of course, to the scale invariance of the gauge theory).
We have seen that the condition for the boundary interaction to prteserve
this one-loop scale invariance is given by (\ref{scal}).
This actually corresponds to the condition that $L$ is
a special Lagrangian submanifold.\footnote{The numbers $m_i^{\alpha}$
are related to ``charges'' $q_i^A$ ($A=1,\ldots,N-\ell$) in \cite{AV}
by the relation $\sum_{i=1}^Nm_i^{\alpha}q_i^A=0$.
Then, the scale invariance condition (\ref{scal}) is equivalent to
the condition $\sum_{i=1}^Nq_i^A=0$ in \cite{AV} for $L$
to be special Lagrangian.}
To see this we note
that the holomorphic volume form of $X$ is proportional to
$\exp(i\sum_{i=1}^N\varphi_i)$. Under the condition (\ref{scal}),
the phase is a constant
\beq
\sum_{i=1}^N\varphi_i=\sum_{i=1}^N\sum_{\alpha=1}^{\ell}
\delta_{\alpha}m_i^{\alpha}\varphi_i
=\sum_{\alpha=1}^{\ell}\delta_{\alpha}{\rm const}_{\alpha}.
\eeq

One important thing to notice here is that
$\varphi_i={\rm Im}\log \phi_i$
is well-defined only if $|\phi_i|^2\ne 0$ but
the equation (\ref{aff}) allows some of $|\phi_i|^2$ to vanish.
Generically, the subspace is singular or has a boundary
at such a point.
It is only in a special case where
(\ref{aff})-(\ref{toru}) defines a smooth submanifold.
For example, let us consider our basic example of a single chiral
superfield $\Phi$ (where there is no gauge symmetry)
and promote $S$ to a boundary chiral superfield.
This will yields the D-brane at $\varphi={\rm Im}\log\phi=$ constant.
This indeed has an end point at $\phi=0$ and is not smooth.
Similarly, in many cases $L$ is singular for any values of $c_i$.
However, there are some cases where $L$ is smooth for special values
of $c_i$.
In such a case, the condition that $L$ is smooth can be considered
as a constraint on the parameters $c_i$.
We exhibit this in the examples below.

\subsubsection*{\it Special Lagrangian Families in $\C^N$}

\begin{figure}[htb]
\begin{center}
\epsfxsize=2in\leavevmode\epsfbox{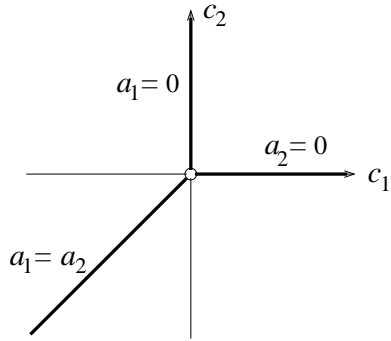}
\end{center}
\caption{The constraint: {\small The three bold open lines
are the locus where $L$ is smooth.
 We have set $a_3=c_3=0$ to eliminate the shift ambiguity.}}
\label{class}
\end{figure}

In this example, we do not consider a gauge theory but a
free theory of $N$ chiral superfields $\Phi_i$.
One can straightforwardly apply the above contruction of
boundary interaction to this case. (Simply ignore
$V$ and $t$ and omit the constraints such as
$\sum_{i=1}^NQ_iS_i=t$.)
In particular, 
(\ref{scal}) is still the condition of one-loop scale invariance.
We will focus on such a case with $\ell=1$.
Namely, the case where $m_i$ are all equal, say to $1$.
The equation defining the subspace $L$ is
\beq
\begin{array}{l}
|\phi_i|^2=\zeta+c_i~~~(i=1,\ldots,N),\\
\varphi_1+\cdots+\varphi_N=0.
\end{array}
\label{defq}
\eeq
This is smooth if and only if one can find a pair $(i_1,i_2)$
such that $c_{i_1}=c_{i_2}< c_j$ for $j\ne i_1,i_2$.
Otherwise, $L$ has an end at the locus where only one $\phi_i$ vanishes.
If the condition holds, say for $(i_1,i_2)=(1,2)$,
$\phi_{1}$ and $\phi_{2}$ vanishes at the same time,
and $L$ is smooth since
the defining equation can be written as
\beqa
&&\bphi_1=\phi_2\times\exp\left(i\mbox{$\sum_{j=3}^N\varphi_j$}\right),
\nn\\
&&|\phi_j|^2=|\phi_1|^2+c_j-c_1,~~(j\ne 1,2).
\nn
\eeqa
In this case, the circle of $\varphi_1$ and $\varphi_2$ are topologically
trivial in $L$ and thus the holonomy must be trivial;
$a_1=a_2=0$. This is relaxed to $a_1=a_2$
by using the freedom to shift $a_i$ uniformly (coming from
the second equation of (\ref{defq})).
In general, we must have $a_i=a_j$ in the branch where $c_i=c_j$.
The space of $(c_i)$'s satisfying the constraint
is a union of walls in the $(N-1)$-dimensional space ($-1$ is from
the redefinition of $\zeta$).
We depict in Fig.~\ref{class}
the case of $N=3$.
The origin is deleted since $L$ is singular there.

\subsubsection*{\it ${\cal O}(-1)\oplus {\cal O}(-1)$ over $\CP^1$}

We next consider the $U(1)$ gauge theory with four matters of charge
$Q_i=1,1,-1,-1$.
We stay in the region where $r$ is large positive where the bulk theory
reduces to non-compact Calabi-Yau sigma model.
We consider the case $\ell=1$ where $m_i=1,1,1,1$.
This is again the case where (\ref{scal}) is met and
the boundary interaction is scale invariant at the one-loop level.
For the gauge invariance and supersymmetry
$c_i$ and $a_i$ are constrained by
$c_1+c_2-c_3-c_4=r$, $a_1+a_2-a_3-a_4=\theta$.
The overall shifts of $c_i$ and $a_i$ are unphysical.
Thus space of physical $c_i$ is two-dimensional
and so is that of $a_i$.
The equation defining the subspace $L$ is
the same as the $N=4$ case of (\ref{defq}).
$L$ is smooth if and only if one can find a pair $(i,j)$
such that $c_i=c_j< c_k$ for $k\ne i,j$.
In the $c_i=c_j$ branch, $a_i=a_j$ must be satisfied.
Thus, the parameter space consists of five disconnected pieces, each being
an open cylinder (four of them semi-infinite).
There are two points where three of the branches becomes close as in
Fig.~\ref{class}.

In the two examples, we have seen that the smoothness of the
submanifold $L$ put a constraint of the parameters $c_i$ and $a_i$.
However, we notice that the geometry itself is derived from the
action involving the boundary D-terms.
We know that boundary D-terms are not protected even from
loop corrections. Accordingly, we expect that the constraint
on $c_i$ and $a_i$ is subject to quantum corrections as well.
We will see that it is indeed the case.

\subsection{The Mirror Description}\label{subsec:mir}

In \cite{HV}, a dual description of the linear sigma model
was found. Dualizing on the phase of each chiral superfield $\Phi_i$
we obtain a twisted chiral superfield $Y_i$
whose real part is related to the gauge invariant composite of
$\Phi_i$ via
\beq
\bPhi_i\e^{Q_iV}\Phi_i={\rm Re}\,Y_i.
\label{dualrel}
\eeq
The dual theory has the (twisted) superpotential
\beq
\widetilde{W}=\Sigma\left(\sum_{i=1}^NQ_iY_i-t\right)
+\sum_{i=1}^N\e^{-Y_i},
\label{dualW}
\eeq
where $\Sigma$-linear term appears at the dualization process and
the exponential terms are generated by the effect
of the instantons which are vortices in the gauge theory.
In the sigma model limit where $e\to\infty$,
it is appropriate to integrate out the gauge multiplet
and that induces the constraint
\beq
\sum_{i=1}^NQ_iY_i=t.
\eeq
The theory becomes the Landau-Ginzburg model
on this $N-1$ dimensional algebraic torus $(\C^{\times})^{N-1}$
with the superpotential
\beq
\widetilde{W}=\sum_{i=1}^N\e^{-Y_i}.
\label{tlW}
\eeq
We would now like to ask how the A-type D-brane constructed above
is described in the dual theory.

We first give a rough argument which in the end turns out
to be the correct one.
Let us look at the boundary interaction term (\ref{Bount}).
Here we replace $s_i$ by $S_i$
which can either be a parameter or a boundary chiral superfield.
If we use the relation (\ref{dualrel}), the
$U_i$ term in (\ref{Bount}) can be made into
a boundary F-term and the total
boundary F-term is expressed as
\beq
{1\over 2\pi}\sum_{i=1}^N
\int\limits_{\partial\Sigma}\dd x^0\,{\rm Re}\!\int \dd\theta
\, (S_i-Y_i)
\Upsilon_i.
\label{redBount}
\eeq
Thus, $\Upsilon_i$ integration yields the constraint
\beq
Y_i=S_i.
\label{Acon}
\eeq
This argument was not precise
for two reasons:
First, the relation (\ref{dualrel}) obtained
in the bulk theory is used without paying attention to the
presence of the boundary.
Second, it ignores the other boundary interactions ---
the boundary term in $S_A$ (see eqn.~(\ref{Sp}))
and the term involving ${\rm Im}\log \Phi_i$ in (\ref{Bount}).
We now show that the result (\ref{redBount}) or (\ref{Acon})
remains correct (with a different interpretation of $\Upsilon_i$)
even if we take these points into account.

We first note that during the dualization procedure
we take $|\phi_i|^2$ to be non-zero and 
$\varphi_i={\rm Im}\log\phi_i$ is well-defined.
This allows us to shift the $U_i$ field
as $U_i=\varphi_i+U'_i$
so that $U_i'$ is a single valued superfield.
In terms of the shifted variables
 the boundary term (\ref{Bount}) is expressed as
\beq
S_{\it boundary}={1\over 2\pi}\sum_{i=1}^N\int\limits_{\partial\Sigma}
\dd x^0
\left[\int\dd\theta\dd\btheta\,
\bPhi_i\e^{Q_iV}\Phi_iU'_i
+{\rm Re}
\int \dd\theta
\,S_i\Upsilon'_i
+a_i\partial_0\varphi_i\,
\right].
\label{Bount2}
\ee
The terms in the action relevant for the dualization
are those involving $\varphi_i$'s:
\beqa
&&-{1\over 2\pi}\sum_{i=1}^N\int\limits_{\partial\Sigma}
|\phi_i|^2(\partial_{\mu}\varphi_i+Q_iv_{\mu})^2\dd^2x
\nn\\
&&~~~~~~~
+{1\over 2\pi}\int\limits_{\partial\Sigma}
\left[\,\,\sum_{i=1}^N\left(
-2u_i'|\phi_i|^2(\partial_1\varphi_i+Q_iv_1)+a_i\partial_0\varphi_i\right)
+\theta v_0\,\,
\right]\dd x^0,~~~
\label{dure}
\eeqa
where $u_i'$ is the lowest component of $U_i'$.
Here we have included the boundary Theta term
${\theta\over 2\pi}\int_{\partial\Sigma}v_0\dd x^0$
(a term in $S_A$) in order to keep the
gauge invariance: Note that (\ref{dure}) itself
is invariant under the gauge
transformation $\varphi_i\to\varphi_i+Q_i\alpha$,
$v_{\mu}\to v_{\mu}-\partial_{\mu}\alpha$ provided
the condition $\sum_{i=1}^NQ_ia_i=\theta$ from (\ref{sgcond1})
holds.
Also we have ignored the terms involving fermions. (This
is merely for simplicity and there is no obstacle to include them
into the discussion below.)
Now we consider a system of $N$ one form fields $(B_i)_{\mu}$ and
$N+1$ periodic scalar fields $\vartheta_i$, $\widetilde{u}$
with the following action
\beq
S'={1\over 2\pi}\sum_{i=1}^N\Biggl[~
\int\limits_{\Sigma}\left(
-{1\over 2}|B_i|^2\dd^2x +B_i\wedge \dd\vartheta_i
+Q_i\vartheta_iF_v\right)
+\int\limits_{\partial\Sigma}
(a_i-\vartheta_i)\partial_0\widetilde{u}\dd x^0
~\Biggr],
\label{Sdu}
\eeq
where $F_v$ is the curvature of $v$, $F_v=\dd v$.
We impose the boundary condition
\beq
(B_i)_1=0.
\label{btan}
\eeq
If we integrate out the $B_i$-fields, we obtain the action for
twisted chiral superfields $Y_i=|\phi_i|^2-i\vartheta_i+\cdots$
and $\Sigma$ with the bulk superpotential
$\widetilde{W}_{\it dual}=\Sigma(\sum_{i=1}^NQ_iY_i-t)$ and
the boundary interaction (\ref{redBount})
in which $\Upsilon_i$ is the ``fieldstrength'' for $U_i$ with the
lowest component $\widetilde{u}_i$.
Thus, the rest is to show that the integration in the opposite order,
$\vartheta_i$ first, yields the action (\ref{dure}).
The variation with respect to $\vartheta_i$ gives the constraints
\beqa
&&\dd B_i=Q_iF_v~~~\mbox{on $\Sigma$}
\nn\\
&&(B_i)_0=\partial_0\widetilde{u}~~~\mbox{along $\partial\Sigma$}.
\nn
\eeqa
The first constraint is solved by $B_i=\dd\varphi_i+Q_iv$
where $\varphi_i$ is a periodic scalar field of period $2\pi$.
By the second equation plus
the boundary condition (\ref{btan})
this leads to the following relations on the boundary
\beqa
&&\partial_0\varphi_i+Q_iv_0=\partial_0\widetilde{u},
\\
&&\partial_1\varphi_i+Q_iv_1=0.
\label{lattercond}
\eeqa
Plugging the first relation back into (\ref{Sdu}) and using the relation
$\sum_{i=1}^NQ_ia_i=\theta$ we obtain the action
(\ref{dure}) without the $u_i'$-dependent terms.
The condition (\ref{lattercond}) is actually equivalent to having
those $u_i'$-dependent terms; integrating out $u_i'$ simply
imposes (\ref{lattercond}).

Dualization is not the end of the story in finding the mirror
description \cite{HV}.
As mensioned above, the bulk superpotential $\sum_{i=1}^N\e^{-Y_i}$
is generated by the instanton effect. Like in that case, one may wonder
if the boundary F-term is generated as well.
We now show that nothing can be generated.
As in \cite{HV}, we extend the gauge symmetry to $U(1)^N$ where each chiral
superfield $\Phi_i$ has charge $1$ under the $i$-th $U(1)$ and neutral
under the others.
We have $N$ FI-Theta parameters $t_i$ which is promoted to a twisted
chiral superfield $T_i$, and $Q_iS_i$ must be the A-boundary value of
$T_i$ for gauge invariance and supersymmetry.
For an appropriate choice of the D-terms
 we have $N$ decoupled systems while for another limit
of the D-term couplings we recover the original system.
Since the deformation of the D-term does not affect the F-terms it is enough
to show that (\ref{redBount}) is not corrected for each $i$.
We first note that $\Upsilon_i$ have boundary R-charge $1$
while $\e^{-Y_i}$ and $\e^{-S_i}$ both have R-charge $2$.
The boundary F-term must be holomorphic in
these superfields, and it must approach the classical
expression (\ref{redBount}) at large $Y_i$ and large $S_i$.
These requirement is satisfied only
by (\ref{redBount}) itself.

\subsection{Quantum Deformation of the Constraint}\label{subsec:QDC}

The mirror of our D-brane is thus given by (\ref{Acon}).
This is true not only when $S_i$ are parameters but also
when they are boundary chiral superfields.
If $S_i$ are parameters, (\ref{Acon}) means that the boundary value of
$Y_i$ is fixed at $s_i$ and we see that the mirror of our D-brane
is a D0-brane at a point in $(\C^{\times})^{N-1}$.
If $S_i$ are functions of $\ell$ boundary chiral superfields
as in (\ref{sifun}), the mirror is a D$(2\ell)$-brane
wrapped on a holomorphic cycle $Z$ defined by
\beq
Y_i=S_i(Z_1,\ldots,Z_{\ell}).
\label{defZ}
\eeq
These are B-type D-branes in the LG model (in the flipped convention where
$Y_i$ are {\it chiral}).
B-type D-branes in LG model were brielfly studied in
\cite{GJS,HIV} and will be studied in some more detail
in Section \ref{sec:BLG}.

One important constraint of worldsheet supersymmetry is that
the bulk superpotential must be a constant on the D-brane.
This itself gives no condition in the case where $S_i$ are parameters since
the mirror D-brane is at a point. 
However, the constraint gives a strong condition
when $S_i$ are functions of the chiral superfield $Z_{\alpha}$
so that the mirror D-brane is wrapped on a cycle $X$ defined
by (\ref{defZ}).
It constrains the functional form to be
\beq
\widetilde{W}=\sum_{i=1}^N\e^{-S_i(Z)}={\rm constant}.
\eeq
This is a very strong constraint and is not satisfied by
a generic function $S_i(Z)$.

Let us see whether this constraint is satisfied in the case of
linear functions
$S_i=\sum_{\alpha=1}^{\ell}m_i^{\alpha}Z_{\alpha}+s_i$.
(This discussion is motivated by \cite{AV} where the same constraint
is obtained from a geometric consideration.)
In this case, the cycle $Z$ is the mirror of our non-compact Lagrangian
subspace $L$.
The condition reads as
\beq
\sum_{i=1}^N\e^{-m_i^{\alpha}Z_{\alpha}-s_i}={\rm constant}.
\eeq
This requires the following condition.
Let us separate the set of $i$'s, $I=\{i\}$,
into groups $I=\cup_a I_a$ where $i$'s in each group $I_a$
have the same $m_i^{\alpha}$. Then, the condition is that
$\sum_{i\in I_a}\e^{-s_i}=0$ for each group $I_a$.
In the case where $\ell=1$ and $m_i=1$ for all $i$,
this reads
\beq
\sum_{i=1}^N\e^{-s_i}=0.
\label{qdc1}
\eeq
This replaces the classical constraint that $L$ is smooth.
In fact, in the asymptotic regions where two of $s_i$ are mush smaller than
others, this reduces to the classical
constraint found in the geometric analysis (up to a shift in the
$a$-angle).
Thus, (\ref{qdc1}) can be considered as
the quantum deformation of the classical constraint.
To see this let us consider the $N=4$ case
and send $r$ to infinity where $X$ approaches $\C^3$.
We will foucus on the region where $s_4\sim t\to\infty$.
Then, the constraint becomes
\beq
\e^{-s_1}+\e^{-s_2}+\e^{-s_3}=0.
\label{qmc}
\eeq
It is easy to see that this reduces to the condtraint depicted in
Fig.~\ref{class} as long as one of $c_i={\rm Re}(s_i)$ is large
compared to the other two.
One important point is that the region where the branches meet
was excluded in the classical description but
the distinct branches are smoothly connected in the quantum description.

There is actually a subtlety
associated with the $a$-angles.
Let us focus on the region $c_1, c_2\ll c_3$.
In the geometric discussion we have found that $c_1=c_2$
and $a_1=a_2$.
On the other habd, the condition (\ref{qmc}) is approximately
$\e^{-c_1+ia_1}+\e^{-c_2+ia_2}=0$ in this asymptotic region.
This means that $c_1=c_2$ but $a_1=a_2+\pi$.
This should be understood in the
original linear sigma model.
Including this point, many things has to be clarified.
For instance, it is important to understand
whether there is a singularity in the parameter space.
This requires a more careful dynamical study of this gauge system
on the half-plane (or the strip).
Also, as computed in \cite{AV} in the geometric model
related to this LG model in the weak sense of \cite{HV},
the space-time superpotential as a function of $s_i$'s
is generated. This means that a non-perturbative correction
to the beta function for $s_i$ is non-zero.
How it is computed directly in the LG model should also be clarified.
We hope to discuss thoes points elsewhere.

\subsubsection*{\it The Mirror of the Compact Torus $T$}

Let us consider the case where $S_i$ are parameters $s_i$
and the original A-type D-brane is wrapped on a compact torus $T$.
In this case, the constraint of $Y_i$ is
\beq
Y_i=s_i.
\eeq
Namely, the mirror D-brane is the D0-brane at a point $(s_i)$
in $(\C^{\times})^{N-1}$.
Then, the condition that the superpotential is a constant on the brane
 is satisfied for any value of $s_i$.
However, as we will discuss in Section~\ref{sec:BLG},
there is a further constraint that
$(s_i)$ must be at the critical point of $\widetilde{W}$.
In the case where $X$ is a compact toric manifold of
positive first Chern class,
one can find such a critical point as many as $\chi(X)$, the Euler number of
$X$. Thus, we find $\chi(X)$ D-branes.
For example, for $X=\CP^1=S^2$, the mirror theory is the
${\cal N}=2$ sine-Gordon model with the superpotential
$\widetilde{W}=\e^{-Y}+q\e^Y$ where $q=\e^{-t}$.
The critical points are $\e^{-Y}=\pm\sqrt{q}$.
Thus, we have two kinds of D-branes with
\beq
\e^{-s_1}=\e^{-s_2}=\pm\sqrt{q}.
\eeq
This means that $c_1=c_2=r/2$ and
$a_1=a_2=0$ or $\pi$.
This corresponds to the D-branes at the equator of $S^2$
with the holonomy $\pm 1$.

In the case where $X$ is a non-compact Calabi-Yau manifold,
$\widetilde{W}$ is of Liouville type and
there is no critical point at finite $Y_i$. Therefore we cannot find
a supersymmetric D0-brane.

\section{B-type D-Branes and Tachyon Condensation}\label{sec:BTachy}

We turn to D-branes which preserve B-type supersymmetry.
We set the phase trivial $\e^{i\beta}=1$ unless otherwise stated.
One of the basic examples is the space filling D-brane
which is described in the non-linear sigma model
(with the trivial $B$-field)
by the full Neumann boundary condition for the bosonic
fields, or more completely by
\beq
D_+\Phi^i=D_-\Phi^i~~~\mbox{at B-boundary},
\eeq
where $\Phi^i$ are the chiral superfields representing the complex
coordinates of the target space. We would like to study more non-trivial
examples in what follows.

\subsection{The System of a D-Brane and an Anti-D-Brane}

The first example we consider is the
D0-brane in the complex plane.
As before we realize the supersymmetric sigma model on the complex plane
by the theory of a single chiral superfield
$\Phi=\phi+\theta^{\alpha}\psi_{\alpha}+\theta^+\theta^-F+\cdots$.
We use the following action
\beqa
S_B&=&{1\over 2\pi}\int\limits_{\Sigma} \dd^2x \left(\,
|\partial_{0}\phi|^2-|\partial_1\phi|^2
+{i\over 2}\opsi_-(\lrd_{\!\!\!0}+\lrd_{\!\!\!1})\psi_-
+{i\over 2}\opsi_+(\lrd_{\!\!\!0}-\lrd_{\!\!\!1})\psi_+
+|F|^2\,\right)
\nn\\
&&+{i\over 4\pi}\int\limits_{\partial\Sigma}
\dd x^0\,\left(\opsi_-\psi_+-\opsi_+\psi_-\right),
\label{cact7}
\eeqa
which is invariant under B-type supersymmetry without
using equation of motion nor any boundary condition.
In the standard approach, the D0-brane
at $\phi=\phi_0$ is described by the supersymmetric Dirichlet boundary
condition for the fields which is conveniently summarized as
\beq
\Phi=\phi_0~~~\mbox{at B-boundary}.
\eeq
In the ``linear model approach''
the same D-brane can be represented by the theory involving a
boundary Fermi superfield ${\Gamma}$ and the boundary interaction
\beq
S_{\it boundary}={1\over 2\pi}
\int_{\partial\Sigma} \dd x^0\,{\rm Re}\int \dd\theta\,
{\Gamma}(\Phi-\phi_0),
\label{sbo}
\eeq
where of course the integration is along the B-type boundary
$\theta^+=\theta^-$. In this way of writing, 
it is manifest that $\phi_0$ is the boundary chiral parameter.
In what follows, we show that this latter formulation appears
very naturally as the infra-red limit of the system of
D2 and anti-D2 branes with a specific tachyon configuration.
Basically, $\Phi-\phi_0$ that appears in (\ref{sbo})
is the tachyon configuration.

\subsubsection{The ${\cal N}=1$ Boundary Interaction}

We consider the system of a D-brane and an anti-D-brane filling the
target space (the complex plane in the above example).
As is well-known e.g. \cite{SenLec},
the low lying spectrum of this system
consists of four parts --- gauge fields $A^1$ and $A^2$ from $p$-$p$
and $\bar p$-$\bar p$ strings\footnote{We refer to D$p$ and
anti-D$p$ branes
as $p$ and $\bar p$ respectively. ``A $p$-$p$ string'' for example
stands for an open string stretched between D$p$ and D$p$.}
and tachyon fields $T$ and $\overline{T}$
from $p$-$\bar p$ and $\bar p$-$p$ strings ---
which constitute the Chan Paton matrix
\beq
\left(\begin{array}{cc}
A^1&T\\
\overline{T}&A^2
\end{array}\right).
\label{CPmat}
\eeq
Thus, the worldsheet path-integral receives a factor
\beq
{\rm Tr}\,P\exp\left[-\oint\limits_{C}\left\{
iA^1_{\tau}\mbox{\footnotesize $\left(\begin{array}{cc}1&0\\
0&0\end{array}\right)$}
+iA^2_{\tau}\mbox{\footnotesize $\left(\begin{array}{cc}0&0\\
0&1\end{array}\right)$}
+T_{(0)}\mbox{\footnotesize $\left(\begin{array}{cc}0&1\\
0&0\end{array}\right)$}
+\overline{T}_{(0)}\mbox{\footnotesize $\left(\begin{array}{cc}0&0\\
1&0\end{array}\right)$}
\right\}\dd\tau\right]
\label{CP}
\eeq
from each boundary component $C$ with coordinate $\tau$.
Here $A^i_{\tau}\dd\tau$ is the pull-back of the
gauge field $A^i$ to the worldline $C$,
and $T_{(0)}$ is the tachyon vertex operator in the
zero picture, $T_{(0)}=\psi^ID_IT$
in which $D_IT=(\partial_I+iA^1_{I}-iA^2_{I})T$
and $\psi^I=\psi^I_++\psi^I_-$.
There is a convenient representation of the Chan-Paton factor using
the complex Clifford algebra \cite{bdryfermi}.
The algebra is generated by $\eta$ and
$\bareta$ obeying
\beq
\{\eta,\bareta\}=1,~~\eta^2=\bareta^2=0,
\eeq
and has the spinor representation spanned by vectors $|0\rangle$,
$\bareta |0\rangle$ (where $|0\rangle$ is annihilated by $\eta$)
on which Chan-Paton matrices are realized as
\beq
\mbox{$\left(\begin{array}{cc}1&0\\
0&0\end{array}\right)$}=\eta\bareta,~~
\mbox{$\left(\begin{array}{cc}0&0\\
0&1\end{array}\right)$}=\bareta\eta,~~
\mbox{$\left(\begin{array}{cc}0&1\\
0&0\end{array}\right)$}=\eta,~~
\mbox{$\left(\begin{array}{cc}0&0\\
1&0\end{array}\right)$}=\bareta.
\eeq
Then, the Chan-Paton factor (\ref{CP}) can be considered as the
partition function of the quantum mechanics represented on
the ($|0\rangle$, $\bareta |0\rangle$) space with the Hamiltonian
\beq
H=iA^1_{\tau}\eta\bareta+iA^2_{\tau}\bareta\eta+T_{(0)}\eta
+\bareta\,\overline{T}_{(0)}.
\label{Ham}
\eeq
Therefore, it has the path-integral representation
\beq
\int{\cal D}\eta{\cal D}\bareta
\,\,\exp\left(-\oint\limits_C\left[\bareta {\cal D}_{\tau}\eta
+T_{(0)}\eta+\bareta \,\overline{T}_{(0)}\right]\dd\tau\,\right),
\label{CP2}
\eeq
where ${\cal D}_{\tau}=\dd/\dd\tau-iA^1_{\tau}+iA^2_{\tau}$.
Note that
the gauge transformation $A^1\to A^1-\dd (\arg g_1)$,
$A^2\to A^2-\dd (\arg g_2)$, $T\to g_1Tg_2^{-1}$,
and $\overline{T}\to g_2\overline{T}g_1^{-1}$
is compasated by the transformation
\beq
\eta\to g_2\eta g_1^{-1},~~\bareta\to g_1\bareta g_2^{-1}.
\eeq
In other words, this can be considered as the
gauge transformation property of the boundary
fields $\eta$ and $\bareta$.

The supersymmetric completion of this system
can be described by introducing the ${\cal N}=1$
boundary superfields including the tachyon $T$ and
the boundary fermion $\eta,\bareta$.
(We come back to the Minkowski signature.)
The former is a bosonic complex boundary superfield
${\bf T}$ which has an expansion
\beq
{\bf T}=T+i\theta_1\psi^ID_IT.
\label{tach}
\eeq
The latter is a fermionic superfield
\beq
{\bf\Gamma}=\eta+i\theta_1G.
\eeq
The boundary interaction that completes the one that appears in
(\ref{CP2}) is given by
\beq
S_{\it boundary}
={1\over 2\pi}\int\limits_{\partial\Sigma} \dd x^0\dd\theta_1\left(
\overline{\bf\Gamma}{\cal D}^1{\bf\Gamma}
+{\bf\Gamma} {\bf T}+\overline{\bf T}\overline{\bf\Gamma}\right).
\label{SSbo}
\eeq
In the above expression, ${\cal D}{\bf \Gamma}$
is defined by
${\cal D}^1{\bf \Gamma}=\left(-i{\partial\over\partial\theta_1}
-2\theta_1(\partial_0-iA_0)\right){\bf\Gamma}$ with
\beq
A_0:=A_I\partial_0\phi^I
-{i\over 4}F_{IJ}\psi^I\psi^J,
\label{gaugf}
\eeq
where $A_I=A^1_{I}-A^2_{I}$
is the gauge field of the relative gauge group $U(1)^{\it rel}$
and $F_{IJ}$ is its curvature
$\partial_IA_J-\partial_JA_I$.
This boundary interaction is invariant under the $U(1)^{\it rel}$
gauge symmetry
\beqa
&&A_I\to A_I-\partial_I\alpha(\phi),
\nn\\
&&{\bf\Gamma}\to\e^{-i\alpha(\phi)}{\bf\Gamma},\\
&&{\bf T}\to\e^{i\alpha(\phi)}{\bf T}.
\nn
\eeqa
It is also invariant under the gauge-modified
supersymmetry transformation
\beq
{\cal Q}^1=-i{\partial\over \partial\theta_1}
+2\theta_1(\partial_0+iqA_0)-iqA_I\psi^I,
\eeq
where $q$ is the $U(1)^{\it rel}$ charge of the field
on which ${\cal Q}^1$ is acting.
For example, ${\bf \Gamma}$ has $q=-1$,
${\bf T}$ has $q=1$, while ${\bf\Phi}^I=\phi^I+i\theta_1\psi^I$
(the restriction of the bulk superfield on the boundary)
has $q=0$.
One can check that the superfield ${\bf T}$, which is a function
of the fields ${\bf \Phi}^I$, has the right transformation property.

\medskip

\noindent
{\bf Remark 1.}
In the above argument we have assumed that the path-integral
(\ref{CP2}) leads to the Hamiltonian (\ref{Ham}).
However, there is a standard operator ordering ambiguity
that is fixed by an explicit regularization scheme;
the first two terms in (\ref{Ham}) could be replaced by
$i(A^1_{\tau}-A^2_{\tau})\eta\bareta$ or
$i(A^2_{\tau}-A^1_{\tau})\bareta\eta$, or a combination
of them.
This ambiguity would be annoying in the following discussion where
 we start with the Lagrangian.
In what follows, instead of the above choice,
we take the ordering where the action (\ref{SSbo})
corresponds to $A^2=0$ and $A^1=A$.
In this ordering, if we would like to have a non-zero $A^2$,
we need to add the term
\beq
{\mit\Delta}S_{\it boundary}=-\int\limits_{\partial\Sigma}
\dd x^0
\left(A^2_I\partial_0\phi^I-{i\over 4}F^2_{IJ}\psi^I\psi^J\right)
\label{delSbo}
\eeq
to (\ref{SSbo}) in which $A=A^1-A^2$.

\noindent
{\bf Remark 2.} 
From the above result, one can also obtain
the boundary interaction for the non-BPS D-branes \cite{nonBPS,horava}
in Type II string theory.
In fact the latter is defined as
the $(-1)^{F_L}$ orbifold of the brane-anti-brane system,
where the orbifolding yields the reality constraint on
${\bf\Gamma}$ and ${\bf T}$
and also projects out the relative gauge field $A_I$.
The resulting boundary interaction
is nothing but the one used in \cite{KHM,KMM2,Justin}.
The real boundary fermion $\eta$
was originally introduced in \cite{DK}
to reproduce the interaction rule of \cite{Dint}
for Type I D0-brane \cite{TypeID0}.

\subsubsection{The Condition of ${\cal N}=2$ Supersymmetry}

We would now like to find the ${\cal N}=2$ extension
of the above result.
Thus, we consider a supersymmetric sigma model on a Kahler manifold
$X$ formulated on a strip $\Sigma=\R\times [0,\pi]$.
On the B-boundary,
the chiral superfields $\Phi^i$ representing the complex coordinates of
$X$ become boundary chiral superfields that are expanded as
\beq
{\Phi}^i=\phi^i+\theta\psi^i-i\theta\btheta\partial_0\phi^i,
\eeq
where $\psi^i=\psi_+^i+\psi_-^i$.
The anti-chiral superfields $\bPhi^{\bi}$
become boundary anti-chiral superfields.

\subsubsection*{\it Boundary Gauge Symmetry}

We start with introducing boundary chiral gauge symmetry.
Let
$\Xi=\xi+\theta J-i\theta\btheta\partial_0\xi$
be a boundary Fermi superfield.
We would like to construct a supersymmetric boundary Lagrangian
that is invariant under the gauge transformation
\beq
\Xi\to\e^{iqA}\Xi,
\eeq
where $A$ is a boundary chiral superfield and $q$ is the charge
of $\Xi$. As in the bulk,
it is appropriate to introduce a real boundary superfield $V_b$
which transforms as
\beq
V_b\to V_b-iA+i\overline{A}.
\eeq
Then a gauge invariant and supersymmetric Lagrangian is given by
\beq
L={1\over 2}\int\dd\theta\dd\btheta\,
\,\overline{\Xi}\e^{qV_b}\Xi.
\label{La}
\eeq
One can choose a ``Wess-Zumino gauge'' where $V_b$ has only
the highest component
\beq
V_b=2\theta\btheta A_0.
\label{WZG}
\eeq
The residual gauge symmetry in the Wess-Zumino gauge
is the one with $A=\alpha-i\theta\btheta\partial_0\alpha$ with real
valued $\alpha$ which acts on the component fields as
\beqa
&&\xi\to\e^{iq\alpha}\xi,~~J\to\e^{iq\alpha}J,\nn\\
&&A_0\to A_0-\partial_0\alpha.\nn
\eeqa
In this gauge, the Lagrangian (\ref{La}) is expressed as
\beq
L={i\over 2}\overline{\xi}{\cal D}_0\xi
-{i\over 2}{\cal D}_0\overline{\xi}\xi
+{1\over 2}|J|^2,
\eeq
where ${\cal D}_0\xi=(\partial_0+iqA_0)\xi$.
The ordinary supersymmetry transformation
$\delta=\epsilon\cQ-\bepsilon\bcQ$ does not preserve the
Wess-Zumino gauge. To find the supersymmetry transformation
of the component fields $\xi, J,A_0$, we must modify it with a gauge
transformation. It turns out that the required gauge transformation is
the one with $iA=2\theta \bepsilon A_0$ and we find
\beqa
&&\delta_{\rm tot}\xi=\epsilon J+iq\alpha_*\xi,\\
&&\delta_{\rm tot}J=-2i\bepsilon {\cal D}_0\xi+iq\alpha_* J,\\
&&\delta_{\rm tot}A_0=-\partial_0\alpha_*,
\label{susyA0}
\eeqa
where $\alpha_*$ comes from an ambiguity in the choice of $iA$.

\subsubsection*{\it Gauge Field and Tachyon on the Brane}

We would like to embed the above construction of gauge invariant
interaction to the system of a D-brane and an anti-D-brane.
We extend the ${\cal N}=1$ superfield ${\bf \Gamma}$
to a boundary Fermi superfield $\Gamma$
by replacing $i\theta_1$ by $\theta$ and adding the top component as
\beq
\Gamma=\eta+\theta G-i\theta\btheta \partial_0\eta.
\eeq
We assign gauge charge $q=-1$ to $\Gamma$.
Here, unlike in the above construction, the boundary gauge symmetry is
linked to the gauge symmetry on the branes. That is, 
the field $A_0$ in (\ref{WZG}) is a function of 
$\phi^I$ and $\psi^I$ defined in (\ref{gaugf}) and 
the gauge transformation parameter $\alpha$
is also considered as a function of $\phi^I$.
There is also a tachyon field (\ref{tach})
that is a function of $\phi^I$ and $\psi^I$. 
The supersymmetry transformation of these fields
are dictated by that of ${\Phi}^i$ and $\overline{\Phi}^{\bi}$,
which are in components given by
\beqa
&&\delta\phi^i=\epsilon\psi^i,~~
\delta\psi^i=-2i\bepsilon\partial_0\phi^i,\\
&&\delta\bphi^{\bi}=-\bepsilon\opsi^{\bi},~~
\delta\opsi^{\bi}=2i\epsilon\partial_0\bphi^{\bi}.
\eeqa
It is easy to see that the transformation of $A_0$ takes the form
(\ref{susyA0}) if and only if
\beq
F_{ij}=F_{\bi\bj}=0.
\label{holcond}
\eeq
Namely if and only if the operator
$\overline{\partial}_A=\dd\bz^{\bj}D_{\bj}$
is nilpotent and defines a holomorphic structure
on the associated complex line bundle.
The gauge transformation parameter $\alpha_*$ that appears in
(\ref{susyA0}) is given by
$\alpha_*=-\epsilon\psi^iA_i+\bepsilon\opsi^{\bj}A_{\bj}$.
The tachyon field ${\bf T}$ has an extension to a boundary chiral
superfield if and only if $T$ is holomorphic,
\beq
D_{\bj}T=0.
\label{holT}
\eeq
The chiral extension is then denoted by
\beq
{\cal T}=T+\theta\psi^iD_iT-i\theta\btheta\partial_0T.
\eeq
One can check that it has the right gauge and supersymmetry
transformation property.
A manifestly ${\cal N}=2$ invariant boundary interaction
that reduces to (\ref{SSbo}) is now
given by
\beq
S_{\it boundary}
={1\over 2\pi}\int\limits_{\partial\Sigma}
\dd x^0\left[~
{1\over 2}\int\dd\theta\dd\btheta\,\,\overline{\Gamma}\e^{-V_b}\Gamma
+{\rm Re}\int\dd\theta \,i\Gamma\,{\cal T}
~\right].
\eeq
In this formulation, $\Gamma$ and ${\cal T}$ have mass dimensions
$0$ and $1/2$ respectively.

Like $A_0$, the supersymmetry transformation of
$A^2_0=A^2_I\partial_0\phi^I-{i\over 4}F^2_{IJ}\psi^I\psi^J$
is a pure gauge if and only if $F^2_{ij}=F^2_{\bi\bj}=0$.
Namely, the latter is the condition of
${\cal N}=2$ supersymmetry of
the anti-brane Wilson line term
(\ref{delSbo}).

To summarize, {\it the system of a D-brane and an anti-D-brane
has (B-type) ${\cal N}=2$ worldsheet supersymmetry if
and only if each of the the gauge fields defines
a holomorphisc structure for the associated line bundle,
and the tachyon field is a holomorphic
section of the relative line bundle.}

\subsubsection{The D0-Branes}

Let us come back to the sigma model on the complex plane $X=\C$.
We consider the following configuration:
\beqa
&&A_{\phi}=A_{\bphi}=0,\\
&&T=\phi-\phi_0.
\eeqa
The tachyon field $T$ is holomorphic with respect to the trivial
gauge connection and this should define an ${\cal N}=2$ supersymmetric
theory. The chiral tachyon superfield is given in this case by
\beq
{\cal T}={\Phi}-\phi_0.
\eeq
The boundary interaction is now
\beq
S_{\it boundary}
={1\over 2\pi}\int\limits_{\partial\Sigma}
\dd x^0\left[~
{1\over 2\lambda^2}\int\dd\theta\dd\btheta\,\,\overline{\Gamma}\Gamma
+{\rm Re}\int\dd\theta \,i\Gamma\,({\Phi}-\phi_0)
~\right],
\eeq
where we have introduced the coupling constant $\lambda$ that has mass
dimension $1/2$ so that ${\cal T}={\Phi}-\phi_0$
has dimension $0$.
The action, including the bulk one (\ref{cact7}),
is at most quadratic in all fields
and the theory is renormalizable by itself.
It is free and all the fields/parameters have their canonical dimension.
In particular, the infra-red limit simply corresponds to $\lambda\to\infty$.
Also the position parameter $\phi_0$ is not renormalized,
as we will show below in a more general context.
In this way, we recover the
interaction (\ref{sbo}) that imposes the constraint
$\Phi=\phi_0$.
Thus, this system is identified as the
D0-brane at $\phi=\phi_0$.

The above is compatible with some knowledge about D0-brane
as a bound state of D2-anti-D2 system.
For instance, it has a unit winding number at infinity
\cite{Tcond,DK}.
For this to be really identified as
the D0-brane, $|T|$ should approach the vacuum value
at infinity $|\phi|\to\infty$.
As proposed/found in \cite{KMM2},
the tachyon potential in this formulation
is given by $\e^{-|T|^2/4}$ and the minimum is
indeed at $|T|=\infty$.
More importantly, following the proposal of \cite{KMM2},
one can compute the open string field theory
effective action as a function of
$\lambda$ (denoted by $u$ in \cite{KMM2}) and it is minimized
indeed at $\lambda\to\infty$.
Furthermore, this computation gives the ratio
of the tensions of the D2-brane and the D0-brane;
We find $T_0/T_2=(2\pi)^2$ which is the correct result (we are taking
the unit $\alpha'=1$: if we recover $\alpha'$ this is
$T_0/T_2=(2\pi)^2\alpha'$).

\subsubsection*{\it The Non-renormalization Theorem}

Let us consider another configurations
\beqa
&&A_{\phi}=A_{\bphi}=0,\\
&&T=P(\phi,a_p)=a_0+a_1\phi+\cdots a_k\phi^k.
\eeqa
This again preserves ${\cal N}=2$ supersymmetry. The chiral tachyon
superfield is given by
\beq
{\cal T}=P(\Phi,a_p),
\eeq
and the boundary interaction is
\beq
S_{\it boundary}
={1\over 2\pi}\int\limits_{\partial\Sigma}
\dd x^0\left[~
{1\over 2\lambda^2}\int\dd\theta\dd\btheta\,\,\overline{\Gamma}\Gamma
+{\rm Re}\int\dd\theta \,i\Gamma P(\Phi,a_p)
~\right].
\eeq
This is no longer quadratic in fields and the system has a
non-trivial interaction.
It may appear hard to controle the quantum correction in this system.
However, supersymmetry strongly constrains quantum corrections to
the boundary F-term.
We note that the system preserves the $U(1)$ R-symmetry 
under which $\Gamma$ and $\Phi$ have charge $1$ and
$0$ respectively.
This shows that the boundary F-term is always linear in $\Gamma$,
and the possible correction resides only in
the boundary superpotential $P(\Phi,a_p)$.
The effective boundary superpotential
must be holomorphic in $\Phi$ and $a_p$'s, and must respect the
$U(1)\times U(1)$ global
``symmetries'' where $\Gamma$, $\Phi$ and $a_p$
have charge $(-1,0)$, $(0,1)$ and $(1,-p)$ respectively.
It is also required to approach the
classical value $P(\Phi,a_p)=\sum_{p=0}^ka_p\Phi^p$
in the limit $a_p\to 0$.
As in \cite{seiberg},
these conditions are enough to constrain
the boundary superpotential not to receive quantum correction at all.
Of course, the boundary D-term $\int\dd\theta\dd\btheta\,{1\over 2\lambda^2}
\overline{\Gamma}\Gamma$ can receive corrections.

\subsubsection*{\it Multiple D0-Branes}

Let us consider the boundary superpotential
\beq
{\cal T}=\prod_{a=1}^k(\Phi-\phi_{a}).
\eeq
We have seen that $\phi_a$ are not renormalized.
When $\phi_{a}\ne \phi_{b}$ for $a\ne b$,
the infra-red limit simply chooses one $\phi_{a}$
and there are in total $k$ copies of the trivial fixed point
we have considered above.
In particular, the boundary entropy at the infra-red limit
is $k$ times that of the trivial one, and so is the tension
\beq
{\rm tension}=kT_0.
\eeq
Thus, this corresponds to $k$ D0-branes located
at $\phi_1,\ldots,\phi_k$.

Let us compute the open string Witten index 
--- ${\rm Tr}(-1)^F$ of the theory formulated on the segment
$0\leq x^1\leq \pi$.
We first consider the case where one end of the string
carries the above boundary interaction and the other end is free
(pure Neumann corresponding to D2-brane).
For the purpose of computing the index, we can take the
zero mode approximattion where we ignore the $x^1$ dependence.
Then, the supercharge $Q$ is given by
\beqa
Q&=&\opsi\dot{\phi}-i\lambda\eta\prod_{a=1}^k(\phi-\phi_a)
\nn\\
&=&-i{\partial\over \partial \bphi}
{\scriptsize \left(\begin{array}{cc}
0&0\\
1&0
\end{array}\right)
\otimes
\left(\begin{array}{cc}
1&0\\
0&1
\end{array}\right)}
-i\lambda\prod_{a=1}^k(\phi-\phi_a)
{\scriptsize \left(\begin{array}{cc}
1&0\\
0&\!\!-1
\end{array}\right)
\otimes
\left(\begin{array}{cc}
0&1\\
0&0
\end{array}\right)},
\eeqa
where the matrix representation is with respect to the basis
$(|0\rangle,\opsi|0\rangle)\otimes (|0\rangle,\bareta|0\rangle)$.
The supersymmetry equation $Q=Q^{\dag}=0$ is solved by the wavefunction
$f|0\rangle+g\opsi\bareta|0\rangle$ where $f$ and $g$ are functions of
$\phi,\bphi$ that obey
\beq
\left(\begin{array}{cc}
-\partial/\partial \bphi&\!\!\lambda\prod_{a=1}^k(\phi-\phi_a)\\
\lambda\prod_{a=1}^k(\bphi-\bphi_a)\!\!&-\partial/\partial \phi
\end{array}\right)
\left(\begin{array}{c}f\\g\end{array}\right)=0.
\eeq
This is identical to the Dirac equation
for the fermion coupled to a $k$-vortex.
As is well known, there are $k$ normalizable solutions for any values of
$\phi_a$.
Thus, we find
that the Witten index is in this case
\beq
{\rm Tr}(-1)^F=k.
\label{ind1}
\eeq
Next let us consider the case where both ends of the string carry
the above boundary interaction.
In that case,
we have two boundary fermions $\eta_0$ and $\eta_{\pi}$,
one at $x^1=0$ and the other at $x^1=\pi$,
and the supercharge $Q$
in the zero mode approximation is given by
\beqa
Q&=&\opsi\dot{\phi}-i\lambda(\eta_0+\eta_{\pi})
\prod_{a=1}^k(\phi-\phi_a)
\nn\\
&=&-i{\partial\over \partial \bphi}\,
{\scriptsize \left(\begin{array}{cc}
\!0\!&\!0\!\\
\!1\!&\!0\!
\end{array}\right)
\!\otimes\!
\left(\begin{array}{cc}
\!1\!&\!0\!\\
\!0\!&\!1\!
\end{array}\right)
\!\otimes\!
\left(\begin{array}{cc}
\!1\!&\!0\!\\
\!0\!&\!1\!
\end{array}\right)
}\nn\\
&&
-i\lambda\prod_{a=1}^k(\phi-\phi_a)
\left\{{\scriptsize \left(\begin{array}{cc}
\!1\!&\!0\!\\
\!0\!&\!\!\!-1\!
\end{array}\right)
\!\otimes\!
\left(\begin{array}{cc}
\!0\!&\!1\!\\
\!0\!&\!0\!
\end{array}\right)
\!\otimes\!
\left(\begin{array}{cc}
\!1\!&\!0\!\\
\!0\!&\!1\!
\end{array}\right)
}
+{\scriptsize \left(\begin{array}{cc}
\!1\!&\!0\!\\
\!0\!&\!\!\!-1\!
\end{array}\right)
\!\otimes\!
\left(\begin{array}{cc}
\!1\!&\!0\!\\
\!0\!&\!\!\!-1\!
\end{array}\right)
\!\otimes\!
\left(\begin{array}{cc}
\!0\!&\!1\!\\
\!0\!&\!0\!
\end{array}\right)
}\right\}
.\nn\\
\eeqa
It is straightforward to show that the number of bosonic and fermionic
supersymmetric ground states are the same. Thus the index in this case
vanishes
\beq
{\rm Tr}(-1)^F=0.
\label{ind2}
\eeq
The results (\ref{ind1}) and (\ref{ind2}) are consistent with
the interpretation of the boundary interaction as $k$ D0-branes;
As is well-known \cite{fiol},
the Witten index in this situation
is the intersection number of the corresponding cycles.
The complex plane $\C$ and $k$ points in $\C$ have
intersection number $k$, whereas the points in $\C$ have
self intersection number zero, in agreement with (\ref{ind1})
and (\ref{ind2}).

\subsubsection*{\it D-Brane wrapped on a Divisor}

The above construction generalizes straightforwardly to
the case where the target space is an arbitrary Kahler manifold $X$.
Let ${\cal L}$ be a holomorphic line bundle over $X$
with a hermitian fibre metric $h$.
We assume that it has a
global holomorphic section $F$.
Let us consider a boundary Fermi-superfield $\Gamma$ with values
in ${\cal L}^{-1}$ and the following boundary interaction
\beq
S_{\it boundary}
={1\over 2\pi}\int\limits_{\partial\Sigma}
\dd x^0\left[~
\int\dd\theta\dd\btheta\,
h^{-1}(\Phi,\bPhi)\overline{\Gamma}\Gamma
+{\rm Re}\int\dd\theta \,i\Gamma F(\Phi)
~\right].
\label{Sbone}
\eeq
This corresponds to a configuration of
the space filling D-brane and anti-D-brane.
The D-brane and the anti-D-brane support the gauge bundle ${\cal L}$
(with the hermitian connection associated with $h$)
and the trivial bundle
${\cal O}_X$ respectively, and
the tachyon configuration is given by
\beq
{\cal O}_X\stackrel{F}{\longrightarrow}{\cal L}.
\eeq
We note that $h(\Phi,\bPhi)$ that appears in the boundary D-term
can receive a lot of quantum corrections, but $F(\Phi)$ is not
renormalized at all.

In the case where the zero of $F$ is simple and $F=0$
is a smooth hypersurface $D$ in $X$, 
we expect that $h(\Phi,\bPhi)$ vanishes in the infra-red limit
and we obtain a constraint
\beq
F(\Phi^i)=0,
\eeq
on the boundary.
Then, the system can be identified as a D-brane wrapped on $D$.
We note that the co-normal bundle (the normal cotangent bundle)
of $D$ in $X$ is equal to ${\cal L}|_D$.
Thus, the Fermi superfield $\Gamma$ can be considered as
taking values in the co-normal bundle of $D$.
This is consistent with the interpretation of
(the lowest component $\eta$ of) $\Gamma$ as the Gamma matrix in the
normal bundle, which is the basic element of the
Atiyah-Bott-Shapiro construction of lower-dimensional D-branes \cite{DK}.
In a more general case where $F=0$ does not define
a smooth submanifold but a divisor $D$, the boundary interaction
may flow to
a non-trivial fixed point. For instance, if $F=f^k$ with $f$ transversal,
the system corresponds to $k$ D-branes at $f=0$.
If $F=f_1f_2$ where $f_1$ and $f_2$ have common zeroes,
it corresponds to intersecting D-branes.

A remark is now in order.
It is natural to expect that the data for the bundle ${\cal L}$ are
chiral parameters.
However, they do not appear in the
boundary superpotential but in the transition function
that relates $\Gamma$'s in different patches.
Thus, in the present description
these parameters are not manifestly chiral.
This is analogous to the similar drawback of the patchwise
description of the non-linear sigma model: it is not
manifest that Kahler class parameters
are twisted chiral.
As we will see shortly, if the bulk theory is realized as
the linear sigma model,
one can find a global descriptions
where patch-wise definition is not necessary.

\subsection{Multiple D-Branes and Anti-D-Branes}

We next consider the system of $m$ D-branes and $\bar m$ anti-D-branes. 
Here we do not construct the boundary interaction
for general configurations,
but provide constructions for a certain class of configurations.
In particular, we consider the case where
$m=\bar m=2^{n-1}$ for some positive integer $n$.
In such a case, the $2^{n-1}+2^{n-1}$ dimensional
Chan-Paton factor is realized on the
irreducible representation $S$ of the $n$-dimensional
complex Clifford algebra
\beq
\{\eta_i,\bareta_i\}=\delta_{i,j},~~
\{\eta_i,\eta_j\}=\{\bareta_i,\bareta_j\}=0,~~~
i,j=1,\ldots,n.
\eeq
The representation $S$ is constructed from a vector 
$|0\rangle$ annihilated by $\eta_i$
by multiplying creation operators $\bareta_i$.
It decomposes into two subspace $S_+$ and $S_-$,
each with dimension $2^{n-1}$, which consist of vectors
$\bareta_{i_1}\bareta_{i_2}\cdots\bareta_{i_s}|0\rangle$
with even $s$ and odd $s$ respectively.
The Chan Paton matrix takes the form (\ref{CPmat})
where the block decomposition corresponds to the decomposition
$S=S_+\oplus S_-$.
The diagonal blocks $A_1,A_2$ are represented by even polynomials
in $\eta_i,\bareta_i$ whereas the off-diagonal blocks $T,\overline{T}$
are represented by odd polynomials.

We consider the sigma model on $\R^{2n}=\C^n$ with the real
coordinates $x^{\mu}$ or the complex coordinates
$\phi^i=x^{2i-1}+ix^{2i}$.
We take the following configuration
\beq
\left(\begin{array}{cc}
A^1&T\\
\overline{T}&A^2
\end{array}\right)
=\sum_{\mu=1}^{2n}x^{\mu}\Gamma^{\mu}
=\sum_{\mu=1}^{2n}x^{\mu}
\left(\begin{array}{cc}
0&\sigma^{\mu}\\
\overline{\sigma}^{\mu}&0
\end{array}\right),
\label{abs}
\eeq
where $\Gamma^{\mu}$ are the $2n$ dimensional Gamma matrices.
This is motivated by the Atiyah-Bott-Shapiro construction \cite{ABS}
that has been proposed to be identified in \cite{DK}
as the tachyon configuration for
the condimension $2n$ D-brane.
(A linear profile is also the
one that is seen from the D-brane probe \cite{khori}.)
Since the Gamma matrices $\Gamma^{\mu}$ are the real and the imaginary parts
of $\eta_i$, this configuration is represented by
\beq
\sum_{i=1}^n\eta_i\phi^i+\sum_{i=1}^n\bareta_i\bphi^i.
\label{abs2}
\eeq
Repeating what we have
done in the system of one D-brane and one anti-D-brane,
we obtain the following boundary interaction corresponding to this
configuration:
\beq
S_{\it boundary}
={1\over 2\pi}\sum_{i=1}^n\int\limits_{\partial\Sigma}
\dd x^0\left[~
{1\over 2\lambda_i^2}\int\dd\theta\dd\btheta\,\,\overline{\Gamma}_i\Gamma_i
+{\rm Re}\int\dd\theta \,i\Gamma_i \Phi^i
~\right].
\label{Sbmulti}
\eeq
Here $\Gamma_i$ are the boundary Fermi superfields
with the lowest component $\eta_i$.

This boundary interaction is quadratic in all fields and
is renormalizable by itself.
In fact, this is simply the sum of $n$ copies of the system
of a D0-brane in the complex plane.
The parameters $\lambda_i$ go to infinity in the
infra-red limit and we obtain the constraint
\beq
\Phi^i=0 ~~~\mbox{at B-boundary}.
\eeq
Thus, this system is identified as a D0-brane
at the origin of $\C^n$.
The partition function of the system is simply given by the
product $\prod_{i=1}^nZ(\lambda_i)$ where
$Z(\lambda_i)$ is the partition function for
the system corresponding to a D0-brane in the complex plane.
Following the proposal of \cite{KMM2} on the
open string field theory action, 
we can compute the ratio of the tensions, say of D9-brane
and D$(9-2n)$-brane.
We find $T_{9-2n}/T_{9}=(2\pi)^{2n}$
which is again the correct result (in $\alpha'=1$).

We note here another representation
of the tachyon configuration (\ref{abs}) \cite{Atiyah,ABS}.
The spinor representation $S$ tensored with the trivial bundle ${\cal O}$
over $\C^n$ can be identified as the exterior algebra over
${\cal O}^{\oplus n}$ under which
\beq
\eta_{i_1}\eta_{i_2}\cdots\eta_{i_k}
|{\rm top}\rangle
\longleftrightarrow
e_{i_1}\wedge e_{i_2}\wedge\cdots\wedge e_{i_k},
\eeq
where
$|{\rm top}\rangle:=\bareta_1\bareta_2\cdots\bareta_n|0\rangle$ and
 $(e_1,\ldots ,e_n)$ is a ``basis'' of ${\cal O}^{\oplus n}$.
The operator $\sum_{i=1}^n\eta_i\phi^i$ that appears in the expression
(\ref{abs2}) is then identified as
the wedge product by $\phi=\sum_{i=1}^ne_i\phi^i$.
From this we see that the tachyon configuration is obtained by folding
the complex
\beq
{\cal O}
\stackrel{\phi\wedge}{\longrightarrow}{\bigwedge}^{\!1\,}
{\cal O}^{\oplus n}
\stackrel{\phi\wedge}{\longrightarrow}
{\bigwedge}^{\!2\,}{\cal O}^{\oplus n}
\stackrel{\phi\wedge}{\longrightarrow}\ldots
\stackrel{\phi\wedge}{\longrightarrow}
{\bigwedge}^{\!n\,}{\cal O}^{\oplus n}
\label{Koszul}
\eeq
into maps between $\bigwedge^{\it even}{\cal O}^{\oplus n}$
and $\bigwedge^{\it odd}{\cal O}^{\oplus n}$.
The complex (\ref{Koszul}) is called the Koszul complex.
We note that the operator $\phi\wedge$ appears in the 
boundary F-term in (\ref{Sbmulti}).

\subsection{D-Branes in Gauge Theory}

Now, we generalize the above construction
of B-type D-branes to supersymmetric gauge theory.
This yields a global description of
D-branes in toric manifolds.\footnote{Boundary conditions
in gauged linear sigma models for B-type D-branes were
studied in \cite{HIV,GJS}.
A construction similar to the one in this subsection
has been presented in the talk \cite{shamit}.}

\subsubsection*{\it Space-filling D-Brane}

We first provide the the boundary interaction corresponding to the
space-filling D-brane in the non-linear sigma model limit.
For simplicity we will mainly be talking about the $U(1)$ gauge theory
with a single chiral matter field, but we will freely
move to more genral cases as the generalization is obvious.
For the bulk action $S$ and other things,
we use the notation fixed in Section~\ref{subsec:gauge}.

The supersymmetry variation of the action $S$ in (\ref{gact})
is a non-vanishing boundary term, as studied in \cite{HIV}.
If we modify the action as
\beqa
S_B&=&S+{1\over 4\pi}\int\limits_{\partial\Sigma}
\dd x^0\,\Biggl[\,
i(\opsi_-\psi_+-\opsi_+\psi_-)-i(\sigma-\bsigma)|\phi|^2
\nn\\[-0.2cm]
&&~~~~~~~~~~~~~~~~~~~+{1\over 2e^2}\left\{\partial_1|\sigma|^2
+2{\rm Im}(\sigma(D+iv_{01}))\right\}
+i(t\sigma-\overline{t}\bsigma)\,\Biggr],~~~~~~~~~
\label{Sb}
\eeqa
it is invariant under 
B-type supersymmetry (with the trivial phase $\e^{i\beta}=1$)
\beq
\delta S_B=0.
\eeq
The boundary condition derived by varying the action $S_B$
contains $\psi_+=\psi_-$ and $\sigma=\bsigma$ which are
completed as
\beq
\begin{array}{l}
{\cal D}_+\Phi={\cal D}_-\Phi,
\\[0.1cm]
~~~~\Sigma=\overline{\Sigma},
\end{array}~~~\mbox{at B-boundary},
\label{bcga}
\eeq
where ${\cal D}_{\pm}=\e^{-V}D_{\pm}\e^V$.
Note that the first condition is gauge invariant
since $\e^V\Phi$ transforms
as $\e^V\Phi\to\e^{i\overline{A}}\e^V\Phi$ with $\overline{A}$ anti-chiral.
We also have another boundary condition
\beq
v_{01}=-e^2\theta.
\eeq
Under the boundary condition (\ref{bcga}),
the boundary terms in (\ref{Sb}) simplifies
so that the action becomes
\beq
S_B=S+{\theta\over 2\pi}\int\limits_{\partial\Sigma}\dd x^0\,
{\sigma+\bsigma\over 2}.
\label{Sb2}
\eeq
In the sigma model limit $e\sqrt{r}\to\infty$,
e.g. in the $\CP^{N-1}$ model where there are $N$ fields of charge $1$,
the field $\sigma$ is frozen at
\beq
\sigma={\sum_{i=1}^N\psi_{i-}\overline{\psi_i}_+\over
\sum_{i=1}^N|\phi|^2}.
\label{sgms}
\eeq
Thus, the boundary term in (\ref{Sb2}) is interpreted as a fermion bilinear
 in the non-linear sigma model.
One can see that this is equal to the fermion bilinear boundary term
in the supersymmetric B-field coupling
\beq
{1\over 2}\int\limits_{\Sigma}B_{IJ}\dd\phi^I\wedge
\dd\phi^J+{i\over 4}\int\limits_{\partial\Sigma}\dd x^0 B_{IJ}\psi^I\psi^J.
\label{Bcoup}
\eeq
Here the bulk B-field term
comes from the Theta term
${\theta\over 2\pi}\int_{\Sigma}v_{01}\dd^2x$
where $v_{\mu}$ is given in the sigma model limit by
\beq
v_{\mu}={i\over 2}{\sum_{i=1}^N\overline{\phi_i}\lrd_{\!\!\!\mu}\phi_i
\over \sum_{i=1}^N|\phi_i|^2}.
\label{vmus}
\eeq
This is a gauge field of the line bundle ${\cal O}(1)$ over $\CP^{N-1}$.

As remarked in \cite{HIV},
there are different formulations
of the non-linear sigma model
when the boundary is coupled to a $U(1)$ gauge field
with non-vanishing field strength.
This is true also when the B-field is non-vanishing.
In one formulation we change the boundary condition of the bosonic fields
from pure Neumann to mixed Dirichlet-Neumann condition, and accordingly
the boundary condition of the fermons is changed as well.
In the other formulation, we do not touch the boundary condition
(for both bosons and fermions)
but, for supersymmetry, we add a fermion-bilinear term on the boundary
as (\ref{Bcoup}).
As explained in \cite{abouel} in the bosonic
string theory, the two formulations lead to the same space-time theory.
The consideration in
\cite{HIV} corresponds to the first formulation.
Here we took the second formulation;
(\ref{bcga}) reduces to
the pure Neumann boundary condition
in the non-linear sigma model limit
and the boundary term in (\ref{Sb2}) reduces to the
boundary term in (\ref{Bcoup}).

\subsubsection*{\it The Boundary Interaction for
Lower-dimensional Branes}

Now we construct a boundary interaction corresponding
 to brane-anti-brane system with tachyon condensation.
To be specific, we consider the $U(1)$ gauge theory with $N$
chiral superfields $\Phi_i$ of charge $Q_i$ which reduces at low enough
energies to the non-linear sigma model on a toric manifold
$X=\C^N/\!/\C^{\times}$.
We denote by ${\cal O}_X(p)$ the line bundle
$(\C^N\times\C)/\!/\C^{\times}$ over $X$
where $\lambda\in \C^{\times}$ acts on the second factor
$c\in\C$ by $c\mapsto \lambda^pc$.

We first recall that the bulk gauge symmetry and the vector superfield
become, when restricted to B-boundary,
a boundary (chiral) gauge symmetry and
a boundary vector superfield.
The bulk Wess-Zumino gauge reduceds to the
Wess-Zumino gauge on the boundary where the vector superfield
is expressed as
\beq
V=2\theta\btheta\left(v_0-{\sigma+\bsigma\over 2}\right).
\eeq
In the sigma model limit (where $v_{\mu}$ and $\sigma$ are given by
(\ref{vmus}) and (\ref{sgms}) in the case
$X=\CP^{N-1}$),
this combination $v_0-(\sigma+\bsigma)/2$ is precisely of the form
$A_0=A_I\partial_0\phi^I-{i\over 4}F_{IJ}\psi^I\psi^J$
that appears in (\ref{gaugf}),
where $A_I$ is the gauge field of the line bundle ${\cal O}_X(1)$.

Let $F(\Phi)$ be a polynomial of $\Phi_i$ of charge $q$.
For this we introduce a boundary Fermi superfield $\Gamma$ of charge
$-q$. We consider the following boundary interaction
\beq
S_{\it boundary}=
{1\over 2\pi}\int\limits_{\partial\Sigma}\dd x^0\left[
{1\over 2 \lambda^2}\int \dd\theta\dd\btheta \,
\overline{\Gamma}\e^{-qV}\Gamma+{\rm Re}\int\dd\theta
\,\Gamma F(\Phi)
\right]
\eeq
This is manifestly gauge invariant and supersymmetric.
At low enough energies where the bulk theory reduces to
the non-linear sigma model on $X$,
 $\Gamma$ reduces to a boundary Fermi superfield
with values in ${\cal O}_X(-q)$ and $F(\Phi)$
determines a holomorphic section of ${\cal O}_X(q)$.
In this limit, the above boundary interaction
reduces to the one given in (\ref{Sbone})
where the hermitian metric $h$ is the
one coming from the standard Euclidean metric of $\C$.
In particular, this boundary interaction corresponds to
the tachyon configuration
\beq
{\cal O}_X\stackrel{F}{\longrightarrow}{\cal O}_X(q).
\eeq
If $F$ has only simple zero, this corresponds to the D-brane wrapped on
the hypersurface $D$
\beq
F=0.
\eeq
Of course, the charge $-q$ for $\Gamma$ is compatible with the fact that
the co-normal bundle of the hypersurface $D$ is equal to
${\cal O}_X(-q)|_D$.

\subsubsection*{\it Intersection of Hypersurfaces}

It is straightforward to generalize the construction
for D-branes wrapped on the intersection
of hypersurfaces $F_1=0,\ldots, F_l=0$.
If $F_{\beta}$ is a charge $q_{\beta}$ polynomial,
the boundary interaction is just
\beq
S_{\it boundary}=\sum_{\beta=1}^l
{1\over 2\pi}\int\limits_{\partial\Sigma}\dd x^0\left[
{1\over 2 e_{b,\beta}^2}\int \dd\theta\dd\btheta \,
\overline{\Gamma}_{\beta}\e^{-q_{\beta}V}\Gamma_{\beta}
+{\rm Re}\int\dd\theta
\,\Gamma_{\beta} F_{\beta}(\Phi)
\right],
\eeq
where $\Gamma_{\beta}$ is a boundary Fermi superfield of charge
$-q_{\beta}$.
We note that the vector R-symmetry (that becomes the boundary R-symmetry
at B-boundary) is always unbroken in the bulk theory.
Thus, the boundary F-term is always linear in $\Gamma_{\beta}$,
and the non-renormalization theorem applies to $F_{\beta}$, as before.

The tachyon configuration is the one obtained from
the Koszul complex associated with
 ${\cal E}=\oplus_{\beta=1}^l{\cal O}_X(q_{\beta})$
and $F=\sum_{\beta=1}^le_{\beta}F_{\beta}$;
\beq
{\cal O}_X
\stackrel{F\wedge}{\longrightarrow}{\bigwedge}^{\!1\,}
{\cal E}
\stackrel{F\wedge}{\longrightarrow}
{\bigwedge}^{\!2\,}{\cal E}
\stackrel{F\wedge}{\longrightarrow}\ldots
\stackrel{F\wedge}{\longrightarrow}
{\bigwedge}^{\!l\,}{\cal E},
\label{Koszul2}
\eeq
by folding into the maps between $\bigwedge^{\it even}{\cal E}$
and $\bigwedge^{\it odd}{\cal E}$.

\section{B-Type D-Branes in Landau-Ginzburg Model}\label{sec:BLG}

In this section, we consider B-type D-branes in a theory
with bulk superpotential.
This leads to the construction of D-branes
in linear sigma models corresponding to hypersurfaces
or complete intersections in toric manifolds.
Another motivation is to study the mirror of the A-type
D-branes identified in Section~\ref{sec:LinA}.

Let us consider a Landau-Ginzburg model
with the superpotential $W(\Phi)$.
The bulk action includes the F-term 
\beq
S_W=\int\limits_{\Sigma}\dd^2x\,{\rm Re}\int\dd^2\theta\, W(\Phi).
\eeq
The B-type supersymmetry transformation of this F-term is
\beq
\delta S_W=\int\limits_{\partial\Sigma}
\dd^2x\,{\rm Re}\int\dd^2\theta\left[
-\bepsilon(\bcQ_++\bcQ_-)W(\Phi)+\epsilon(\cQ_++\cQ_-)W(\Phi)
\right].
\eeq
Using the relations $\cQ_{\pm}W(\Phi)|_{\btheta_{\pm}=0}
={\partial\over \partial\theta^{\pm}}W(\Phi)|_{\btheta^{\pm}=0}$ and
 $\bcQ_{\pm}W(\Phi)
=(\bD_{\pm}-2i\theta^{\pm}\partial_{\pm})W(\Phi)
=-2i\theta^{\pm}\partial_{\pm}W(\Phi)$,
and performaing the partial integration,
we obtain
\beqa
\delta S_W&=&\int\limits_{\partial\Sigma}\dd x^0\,
{\rm Re}\int \dd^2\theta\, i\bepsilon(\theta^+-\theta^-)
W(\Phi)
\nn\\
&=&\int\limits_{\partial\Sigma}\dd x^0\,
{\rm Re}\int\limits_B \dd\theta\, (-i\bepsilon) W(\Phi).
\eeqa
This vanishes if we require
\beq
W(\Phi)={\rm constant}~~~\mbox{at B-boundary}.
\label{constconst}
\eeq
This is the case if we consider a D-brane
on which $W$ is a constant, which
is the condition found in \cite{HIV} by component analysis.

We can apply this argument to the linear sigma model
which corresponds to sigma models on a
complete intersection $M$ in a toric manifold $X$.
In this case, it is natural to choose zero as the constant value of
the superpotential (\ref{constconst}).
Combining this with the construction in the previous
subsection, we can construct the boundary interaction that
corresponds to D-branes in $M$ wrapped on the
holomorphic cycles defined as the intersection of $M$
and  $F_1=\cdots =F_l=0$.

\subsection{D0-Branes in Massive Theory}

In what follows, we focus our attention to
D0-branes in massive LG models.
By {\it massive}, we mean that all the critical points 
of the superpotential $W$ are non-degenerate.
In other words, all the critical points
are isolated and the Hessian (the determinant of
the second derivative matrix) is non-vanishing at each of them.
More general cases such as higher dimensional D-branes
in scale invariant models are also important, say, for string theory
applications, but they will be discussed elsewhere.

Since D0-brane is a point, the condition (\ref{constconst})
is vacuous.
However, if the point is not one of the critical points,
any classical configuration will not attain the zero energy.
We expect that the worldsheet supersymmetry will
be spontaneously broken.\footnote{Non-zero energy does not necessarily mean
supersymmetry breaking, as the example of A-type D-branes
in LG model shows \cite{HIV}.}
To examine this, let us look at the expression
of the supercharge
\beq
Q={1\over 2\pi}\int \dd x^1\left\{
g_{i\bj}(\opsi_-^{\bj}+\opsi_+^{\bj})\partial_0\phi^i
+g_{i\bj}(\opsi_-^{\bj}-\opsi_+^{\bj})\partial_1\phi^i
+(\psi_-^i-\psi_+^i)\partial_iW\right\}.
\eeq
Since the boundary point, say at $x^1=\pi$ is locked at that point,
we see that the supersymmetry is indeed broken for any configuration.
Thus, we will not consider such a D-brane.
In other words, {\it D0-branes must be
located at one of the critical points of $W$}.

\subsection{Supersymmetric Ground States}

Let us compute the supersymmetry index or, if possible,
determine the supersymmetric ground states for an open string
stretched between two critical points $p_a$ and $p_b$ of $W$.
Let us first consider the case $p_a\ne p_b$.
Then, by the same reason as above, there will be
no supersymmetric ground states.
In particular, the supersymmetry index vanishes
\beq
{\rm Tr}_{ab}(-1)^F=0~~~a\ne b.
\eeq
Let us next consider the case where $a=b$.
One can always choose the variables so that $\phi_i=0$ at $p_a$
and
$W=\sum_{i=1}^n m\Phi_i^2+\cdots$ where $+\cdots$ are cubic or
higher order terms.
For the purpose of computing the index, one can deform the Kahler potential
so that it takes the form $K=\sum_{i=1}^n|\Phi_i|^2+\cdots$
and one can also neglect the higher order terms $+\cdots$ in $K$ and $W$.
Then, the computation reduces to that of the free massive theory.

\subsubsection*{\it The Free Massive Theory}

\newcommand{\bm}{\overline{m}}
\newcommand{\bb}{\overline{b}}
\newcommand{\bc}{\overline{c}}
\newcommand{\oxi}{\overline{\xi}}

We are thus led to consider the theory of a single chiral
superfield $\Phi$ with the Kahler potential $|\Phi|^2$ and the
superpotential
\beq
W=m\Phi^2.
\eeq
Since this theory is free, not only the Witten index, but also the
complete spectrum can be determined.

The action of the strip $\R\times [0,\pi]$
after elimination of the auxiliary field is given by
\beqa
S&=&{1\over 2\pi}\int\limits_{\R\times[0,\pi]} \dd^2x\,
\Bigl(~|\partial_0\phi|^2-|\partial_1\phi|^2-|m\phi|^2
+i\opsi_-(\partial_0+\partial_1)\psi_-
+i\opsi_+(\partial_0-\partial_1)\psi_+
\nn\\[-0.2cm]
&&~~~~~~~~~~~~~~~~~~~
-m\psi_+\psi_--\bm\opsi_-\opsi_+~\Bigr).
\eeqa
The boundary condition is that of the D-brane at the critical point
$\phi=0$:
\beq
\Phi=0~~~\mbox{at B-bounday}.
\eeq
In components, this is
\beq
\begin{array}{l}
\phi=0,\\
\psi_-+\psi_-=0,
\end{array}
~~~\mbox{at $x^1=0,\pi$}.
\eeq
The B-type supersymmetry transformation is given by
\beq
\begin{array}{l}
\delta\phi=\epsilon(\psi_-+\psi_+),\\
\delta(\psi_-+\psi_+)=-2i\bepsilon\partial_0\phi,\\
\delta (\psi_--\psi_+)=2i\bepsilon\partial_1\phi+2\epsilon \bm\bphi.
\end{array}
\label{susyv}
\eeq
This motivates us to change the variables as
\beq
b:={\psi_-+\psi_+\over \sqrt{2}},~~~
c:={\psi_--\psi_+\over \sqrt{2}},
\eeq
so that the boundary condition is simply $b=0$ at $x^1=0,\pi$.
In terms of these variables, the fermionic part of
the action can be written as
\beq
S_F={1\over 2\pi}\int\limits_{\R\times [0,\pi]}
\dd^2x\Bigl(~i\bb\partial_0b+i\bc\partial_0 c
+\bb(i\partial_1c+\bm\bc)
+(-i\partial_1\bc+mc)b
~\Bigr)
\eeq
From this we see that we also need to impose the boundary condition
$i\partial_1c+\bm\bc=0$ at $x^1=0,\pi$.
Thus, we are led to the following mode expansion:
\beqa
&&\phi=\sum_{n=1}^{\infty}\phi_n(\e^{inx^1}-\e^{-inx^1}),\\
&&b=\sum_{n=1}^{\infty}b_n(\e^{inx^1}-\e^{-inx^1}),\\
&&i\partial_1c+\bm\bc=\sum_{n=1}^{\infty}
\sqrt{n^2+|m|^2}d_n(\e^{inx^1}-\e^{-inx^1}),
\eeqa
where $\sqrt{n^2+|m|^2}$ is for later convenience.
The last equation and its complex conjugate are solved by
\beq
c=c_0(x)-\sum_{n=1}^{\infty}\left\{
{n\over \sqrt{n^2+|m|^2}}d_n(\e^{inx^1}+\e^{-inx^1})
+{\bm\over \sqrt{n^2+|m|^2}}d^{\dag}_n(\e^{inx^1}-\e^{-inx^1})\right\}.
\eeq
where $c_0(x)$ solves the equations
$i\partial_1c+\bm\bc=0$ and $-i\partial_1\bc+mc=0$.
A general solution of the latter is
given by
\beq
c_0(x)=c_{0}^+\e^{|m|x^1}+c_{0}^-\e^{-|m|x^1}
\label{defc0}
\eeq
in which $c_{0}^{\pm}$ is ``real'' in the sense that
\beq
(c_{0}^{\pm})^{\dag}=\mp i{m\over |m|}c_{0}^{\pm}.
\eeq
In terms of these Fourier modes, the action can be written as
$S=\int \dd t L$ where
\beq
L=i\oxi\dot{\xi}+\sum_{n=1}^{\infty}
\left\{
|\dot{\phi}_n|^2-(n^2+|m|^2)|\phi_n|^2
+ib_n^{\dag}\dot{b}_n
+id_n^{\dag}\dot{d}_n
+\sqrt{n^2+|m|^2}(b_n^{\dag}d_n+d_n^{\dag}b_n)\right\},
\eeq
in which $\xi$ is the following complex combination of
$c_0^+$ and $c_0^-$:
\beq
\xi=\sqrt{\e^{\pi|m|}-\e^{-\pi|m|}\over 4\pi |m|}\Bigl(\xi^++i\xi^-);
~~~
\xi^{\pm}=\e^{\pm \pi|m|/2}\sqrt{\mp i{m\over |m|}}c_0^{\pm}.
\eeq
This system is indeed supersymmetric with respect to
the variation
\beq
\delta\phi_n=\epsilon b_n,~~
\delta b_n=-i\bepsilon \dot{\phi}_n,~~
\delta d_n=\bepsilon \sqrt{n^2+|m|^2}\phi_n,~~
\delta\xi=0,
\eeq
that follows from (\ref{susyv}).
The system of $\phi_n$, $b_n$, $d_n$ for each $n$ is the
(complexified) supersymmetric harmonic oscillator and the quantization
is standard. In particular, it has
a unique supersymmetric ground state $|0\rangle_n$.
On the other hand, the zero mode system of $\xi$ has vanishing Hamiltonian
and the two states $|0\rangle_0$ and $\oxi|0\rangle_0$
are both supersymmetric ground states.
Thus, we see that the total system has two supersymmetric ground states
\beq
|0\rangle,~~\oxi|0\rangle,
\eeq
where $|0\rangle=\otimes_{n=0}^{\infty}|0\rangle_n$.
In particular the index vanishes
\beq
{\rm Tr}(-1)^F=0.
\eeq

\subsubsection*{\it The General Case}

From the above analysis, we conclude in the general massive LG model
that
\beq
{\rm Tr}_{ab}(-1)^F=0~~~\mbox{\it for any $a$ and $b$}.
\eeq
Namely, the index vanishes not only for $a\ne b$ but also for $a=b$.
Moreover, in the case where the quadratic approximation around
the critical point is good enough,
we see from the above analysis that (for $a=b$) there are
$2^n$ supersymmetric ground states if there are $n$ LG fields,
half bosonic and half fermionic;
\beq
|0\rangle,~~\oxi_i|0\rangle,~~\oxi_i\oxi_j|0\rangle,~\ldots,~
\oxi_1\oxi_2\cdots\oxi_n|0\rangle.
\label{susygs}
\eeq
We claim that this is true for {\it any} critical point $p_a$
of a massive LG theory if the manifold on which
the superpotential is defined is Calabi-Yau
(like $\C^n$ or $(\C^{\times})^n$).
Namely, in such a theory, quadratic approximation is
always exact as long as determining the supersymmetric ground states
is concerned.
This can be seen by the correspondence of the supersymmetric
ground states and the boundary chiral ring elements.
To explain this it is best to perform topological twisting.

\subsection{Open Topological Landau-Ginzburg Model}

The twisting can be performed, as usual, by
gauging the $U(1)$ R-symmetry by the worldsheet spin connection.
In the present case, the vector $U(1)$ R-symmetry is broken
by the massive superpotential.
We assume here that the target space $M$ on which
the LG superpotential is defined is a non-compact Calabi-Yau manifold
so that the axial $U(1)$ R-symmetry is unbroken and
we can twist the theory (B-twist).
Twisting changes the spin of the fields as shown by the
new notation below
\beq
\begin{array}{ll}
\opsi_-^{\bi}=\psi^{\bi},&\opsi_+^{\bi}=\opsi^{\bi},\\
\psi_-=\rho_{z}^i,&\psi_+^i=\rho_{\bz}^i,
\end{array}
\label{rename}
\eeq
($\psi,\opsi$ are scalars while $\rho_z$ and $\rho_{\bz}$ define
a 1-form).
Energy-momentum tensor is exact with respect to the operator
\beq
Q=\overline{Q}_++\overline{Q}_-,
\label{BRST}
\eeq
which is scalar after twisting, and we define the space of
``physical operators'' as the $Q$-cohomology of the operators.
The variation of the fields $\delta=-\bepsilon Q$ in the new notation
is given by
\beq
\begin{array}{l}
\delta\phi^i=0,\\
\delta(\psi^{\bi}-\opsi^{\bi})=\bepsilon g^{\bi j}\partial_j W,\\
\delta \rho_{\mu}^i=-2\bepsilon J^{\nu}_{\,\,\mu}\partial_{\nu}\phi^i,
\end{array}
~~
\begin{array}{l}
\delta\bphi^{\bi}=\bepsilon(\psi^{\bi}+\opsi^{\bi}),\\
\delta(\psi^{\bi}+\opsi^{\bi})=0,
\end{array}
\label{delBRST}
\eeq
where $J^{\nu}_{\,\,\mu}$ is the worldsheet complex structure.
From this we see that the physical operators are the holomorphic functions
of $\phi^i$ (i.e. holomorphic functions on $M$) modulo
functions of the form $v^i\partial_i W$ where $v^i\partial_i$
is a holomorphic vector field on $M$.
The physical operators are in one-to-one correspondense
with the supersymmetric ground states of the original LG model
on the periodic circle
(which are identified as the $Q$-cohomology classes of states).
The state corresponding to an operator $O$ is the one that appears at
the boundary circle of the semi-infinite cigar of the twisted model
where $O$ is inserted at the tip \cite{CV}.
The correlation functions of operators $O_1,\ldots,O_s$
on a Riemann surface of genus $g$ is given by
\beq
\langle O_1\cdots O_s\rangle_g=\sum_{p_a:{\rm critical~ point}}
O_1(p_a)\cdots O_s(p_a)(\det\partial_i\partial_j W)^{g-1}(p_a).
\eeq
Here the coordinates defining the derivatives $\partial_i\partial_jW$
are such that the holomorphic $n$-form $\Omega$ is expressed as
$\dd \phi^1\wedge\cdots \wedge \dd\phi^n$ (a choice of $\Omega$ is
required because of the chiral fermionic
determinant).

The above summarizes the topological LG model on a Riemann surface
without a boundary \cite{topLG}.
One important thing to notice is that the operator $Q$ in (\ref{BRST})
is the one that is conserved when the theory is formulated on
the strip with B-type boundary conditions/interactions.
This suggests that one can also consider twisting
B-type boundary theory.
(B-boundary breaks the axial $U(1)_R$ but this is not a problem:
the worldsheetboundary also breaks the local rotation symmetry.)
For D0-branes in a massive LG model,
we only have to translate the boundary condition.
The Wick rotation to the Euclidean signature has to be made first
on the strip \cite{HIV}:
we continue $x^0\to -i x^2$ and the complex coordinate $z$ that
appears in (\ref{rename}) is defined by $z=x^1+ix^2$.
Then, the boundary condition $\phi^i=$ const, $\psi_-^i+\psi_+^i=0$
and $\opsi_-^{\bi}+\opsi_+^{\bi}=0$ is translated as
\beqa
&&\phi^i={\it const},\\
&&\rho^i_n=0,\\
&&\psi^{\bi}+\opsi^{\bi}=0,
\eeqa
where {\it const} is the coordintae value of a critical point, say $p_a$,
and $\rho_n^i$ is the normal component to the boundary.
In particular, the fields remaining on the boundary are
$\theta^{\bi}:=\psi^{\bi}-\opsi^{\bi}$,
the tangent component $\rho_{\tau}^i$ of $\rho^i$,
and the normal derivatives of all fields
including $\eta^{\bi}:=\psi^{\bi}+\opsi^{\bi}$.
The $Q$-variation of these fields can be read from (\ref{delBRST})
as
\beq
\begin{array}{l}
\delta\partial_n\phi^i=0,\\
\delta\theta^{\bi}=0,\\
\delta \rho_{\tau}^i=2\bepsilon \partial_n\phi^i,
\end{array}
~~
\begin{array}{l}
\delta \partial_n\bphi^{\bi}=\bepsilon \partial_n\eta^{\bi},\\
\delta \partial_n\eta^{\bi}=0.
\end{array}
\eeq
From this we see that $Q$-cohomology classes are made of $\theta^{\bi}$
and there are $2^n$ of them:
\beq
1,~~\theta^{\bi},~~\theta^{\bi}\theta^{\bj},~\ldots,
~\theta^{\bar 1}\theta^{\bar 2}\cdots\theta^{\bar n}.
\label{bchiral}
\eeq
As in the case without boundary, there is a one-to-one
correspondence between the supersymmetric ground states of the
original LG model on the segement $[0,\pi]$ 
and the $Q$-cohomology classes of the boundary operators;
The state corresponding to an operator $O$ is the one that appears at
the back of a semi-infinite thin tang of the twisted model
where $O$ is inserted at the tip \cite{BCOV}.
Therefore, we have established from
(\ref{bchiral})
that the spectrum (\ref{susygs}) of the supersymmetric
ground states is an exact result.

Let us compute some correlation functions.
We first consider the correlation functions on the finite size cylinder,
$\Sigma=S^1\times [0,\pi]$.
We impose the boundary condition corresponding to
 the D0-brane at $p_a$ and $p_b$ at the boundary circles
$S^1\times 0$ and $S^1\times \pi$.
If $p_a\ne p_b$, no configuration is $Q$-invariant and therefore
all the topological correlation function vanishes.
If $p_a=p_b$, the constant map to $p_a$ is $Q$-invariant and the
path-integral can be exactly performed by the
quadtratic approximation around the constant map.
Thus, we can compute the correlators using the free massive theory.
(One can consider it as a sum of $n$ decoupled free system;
Since the deformation of D-term does not affect the topological
correlation functions,
one can choose the Kahler potential so that
$K=\sum_{i=1}^n |\Phi_i|^2+\cdots$ at the same time
as $W=\sum_{i=1}^n m_i\Phi_i^2+\cdots$.)
First thing to notice is that there are $2n$ fermionic zero modes
on the cylinder; for each $i$ the functions $c_0(x)$ and $\bc_0(x)$
as in (\ref{defc0}) with $|m|\bc_0^{\pm}=\mp i m c_0^{\pm}$
define the zero mode.
Thus, we must insert $2n$ fermionic operators
for the amplitude to be non-vanishing. 
In particular, the partition function vanishes (this rederives
${\rm Tr}(-1)^F=0$).
Now, let us insert
\beq
(\theta)^n:=\theta^{\bar 1}\theta^{\bar 2}\cdots \theta^{\bar n}
\eeq
at a point of each boundary circle.
The computation reduces to that of the $n=1$ free massive theory
which we have studied above. In fact
we have already developed the machinery of computation.
Since $\theta=\opsi_--\opsi_+=\sqrt{2}\bc$,
what we want to compute is
\beq
{\rm Tr}((-1)^F\e^{-\beta H}2\bc(0)\bc(\pi)).
\label{trcc}
\eeq
It is easy to see that only the zero mode $\bc_0(0)\bc_0(\pi)$
contribute in this computation. 
In terms of the normalized variables $\xi^{\pm}$ or $\xi$, $\oxi$, 
we have
\beqa
\bc_0(0)\bc_0(\pi)&=&
\left(\sqrt{-i{m\over |m|}}\e^{-{\pi|m|\over 2}}\xi^+
+\sqrt{i{m\over|m|}}\e^{{\pi|m|\over 2}}\xi^-\right)
\left(\sqrt{-i{m\over |m|}}\e^{{\pi|m|\over 2}}\xi^+
+\sqrt{i{m\over|m|}}\e^{-{\pi|m|\over 2}}\xi^-\right)
\nn\\
&=&{m \over |m|}(\e^{\pi |m|}-\e^{-\pi |m|})\xi^-\xi^+
=2\pi i\,m\, \oxi\xi.
\eeqa
Then (\ref{trcc}) is
\beq
4\pi im\left(\langle 0|\oxi\xi|0\rangle
-\langle 0|\xi\oxi\xi \oxi|0\rangle\right)=4\pi im(0-1)
=-4\pi im.
\eeq
We note that $m$ is the second derivative of the superpotential
$W''(\phi=0)=2m$.
Thus, the correlation function
in the general case is given by (up to a numerical factor)
\beq
\langle (\theta)^n(0)(\theta)^n(\pi)\rangle^{aa}_{\rm cylinder}
=\det\partial_i\partial_j W(p_a).
\label{cylamp}
\eeq
On the other hand,
by the factorization of the topological correlators,
we have
\beq
\langle (\theta)^n(0)(\theta)^n(\pi)\rangle^{aa}_{\rm cylinder}
=\langle (\theta)^n O_c\rangle^a_{\rm disk}
\eta^{cd}\langle O_d(\theta)^n\rangle^a_{\rm disk},
\label{cylamp2}
\eeq
where $\eta^{ab}$ is the inverse matrix of
$\eta_{ab}=\sum_c O_a(p_c)O_b(p_c)/\det\partial_i\partial_j W(p_c)$.
If we choose as the basis of physical operators the functions
$\epsilon_a$ such that $\epsilon_a(p_b)=\delta_{a,b}$,
then we have $\eta^{ab}=\delta_{a,b}\det\partial_i\partial_j W(p_a)$.
It thus follows from (\ref{cylamp}) and (\ref{cylamp2}) that
$\langle (\theta)^n \epsilon_a\rangle^a_{\rm disk}=1$.
Also it is obvious that
$\langle (\theta)^n \epsilon_b\rangle^a_{\rm disk}=0$ if $b\ne a$.
To summarize, we have obtained
\beq
\langle (\theta)^n O\rangle^a_{\rm disk}=O(p_a).
\label{Opa}
\eeq

\subsubsection*{\it The Sine-Gordon Model}

As an example, let us consider the ${\cal N}=2$ supersymmetric
sine-Gordon model;
the LG model of a cylindrical variable $\e^{-Y}$
with the superpotential
\beq
W=\e^{-Y}+q\e^{Y}.
\eeq
This superpotential has two critical points
$\e^{-Y}=\pm\sqrt{q}$ at which the Hessian is
$\partial_Y^2W=\pm 2\sqrt{q}$.
Some non-vanishing sphere amplitudes are
\beqa
&&\langle \e^{-Y}\rangle_{S^2}=1,
\label{lYr}\\
&&\langle \e^{-Y}\e^{-Y}\e^{-Y}\rangle_{S^2}=q
\label{lYYYr}.
\eeqa
On the other hand, the disk amplitudes are computed using
(\ref{Opa}) as
\beqa
&&\langle \theta \rangle^{\pm}_{\rm disk}=1,
\label{ltr}\\
&&\langle \theta \e^{-Y}\rangle^{\pm}_{\rm disk}=\pm\sqrt{q},
\label{ltYr}
\eeqa
where the superscript $\pm$ stands for the location of the D-brane,
$\e^{-Y}=\pm\sqrt{q}$.

\subsection{Comparison to the Sigma Models}

The sigma model on an $n$-dimensional toric manifold
is mirror to a LG model of $n$ variables.
As shown in Section \ref{subsec:mir}, the D0-branes in such a LG model
are the mirror of the D-branes wrapped on a certain Lagrangian torus.
Thus, such D-branes in the toric sigma model must have the same
properties as D0-branes in the LG model studied in this section
when the theory is massive:
Open string Witten index must be zero for any pair of D-branes;
Space of supersymmetric ground states must be $2^n$ dimensional as in
(\ref{susygs}); the topologically twisted theory
must have the same correlation functions.
Here we check some of these properties directly in the
non-linear sigma model (although it is not necessary because we have
a {\it proof} of the mirror symmetry).

For our purpose, it is convenient to start with twisting the theory.
We are now considering A-twist where $Q=\overline{Q}_++Q_-$
becomes the scalar operator that defines ``physical operators''.
If we put A-type boundary condition/interaction,
we can also consider twisting the theory with boundaries.
We recall that our D-brane is wrapped on a real $n$-dimensional
torus $T$ embedded in the toric manifold $X$.
The theory has $t$, the complexified Kahler class,
and $s_i$, the parameters determing the location of $T$
and the holonomy of flat $U(1)$ bundle on $T$ (which are related so that
$(\e^{-s_i})$ is at the critical point of (\ref{tlW})).
The twisted theory depends only on these parameters and independent
of the detail of the metric.
In particular, one can choose the metric of $X$ so that
a neighborhood of $T$ is that of the $n$-torus $T^n$ in the flat cylinder
$\C^n/\Z^n$, where $T^n$ is the real section $\R^n/\Z^n$.
This is possible at the values of $s_i$ we are choosing.

Now, the A-twist changes the spin of the fermions so that
the following renaming is natural
\beq
\begin{array}{ll}
\psi_-^i=\chi^i,&\opsi_+^{\bi}=\bchi^{\bi},\\
\opsi_-^{\bi}=\rho^{\bi}_z,&\psi_+^i=\rho^i_{\bz}.
\end{array}
\eeq
The boundary condition $\phi^i=\bphi^{\bi}$,
$\psi_-^i=\opsi_+^{\bi}$ and $\opsi_-^{\bi}=\psi_+^i$
is translated to
\beqa
&&\phi^i=\bphi^{\bi},\\
&&\chi^i=\bchi^{\bi},\\
&&\rho^{\bi}_z=\rho_{\bz}^i,
\eeqa
where $z$ is a worldsheet coordinate whose real part is
normal to the boundary.
The variation $\delta=\epsilon Q$ of the remaining variables
at the boundary is given by
\beqa
&&\delta (\phi^i+\bphi^{\bi})=\epsilon(\chi^i+\bchi^{\bi}),\\
&&\delta (\chi^i+\bchi^{\bi})=0,\\
&&\delta(\rho^{\bi}_z+\rho^i_{\bz})
=2i\epsilon(\partial_{\bz}\phi^i-\partial_z\bphi^{\bi}).
\eeqa
From this we see that the $Q$-cohomology classes are in one-to-one
correspondence with the de Rham cohomology classes
of the torus $T^n$, or
\beq
\mbox{$Q$-cohomology group} =
H_{DR}^*(T^n),
\eeq
just as $Q$-cohomology group = $H_{DR}^*(X)$ in the bulk theory.
By the state-operator correspondence as before,
this is identified as
the space of supersymmetric ground states of the theory on the
segment with the same boundary condition at the two ends.
Obviously, $H_{DR}^*(T^n)$ is $2^n$-dimensional and has a basis
like (\ref{susygs}).
In particular, Witten index, identified as the Euler number of $T^n$,
 vanishes.

\subsubsection*{\it The $\CP^1$ Model}

Let us consider the $\CP^1$ model.
This theory is mirror to the sine Gordon model with $q=\e^{-t}$
and must reproduce the result obtained in the previous subsection.
As we have seen in Section~\ref{subsec:QDC},
the D-branes for the two values of $(s_i)$,
$s_1=s_2=t/2$ and $s_1=s_2=t/2+\pi i$, are both wrapped on
the equator of $\CP^1$ but differ in the Wilson line.

The topological $\CP^1$ sigma model for worldsheet without a boundary has
been well-studied.
It has two operators $1$ and $H$ where $H$ is a second cohomology class
represented by a delta-function 2-form at a point of $\CP^1$.
Non-vanishing sphere correlation functions are
\beqa
&&\langle H\rangle_{S^2}=1,\label{lHr}\\
&&\langle HHH\rangle_{S^2}=\e^{-t}\label{lHHHr}.
\eeqa
The first comes from the constant maps; one insertion of $H$
require the insertion point to be mapped a given point of $\CP^1$
and there is only one such map.
The second comes from the degree $1$ maps;
the three insertions of $H$ requires three insertion points
to be mapped to given three points (one for each) of $\CP^1$
and there is one such map. The factor $\e^{-t}$ comes from
the classical action.
The result (\ref{lHr}) and (\ref{lHHHr})
are in agreement with (\ref{lYr}) and (\ref{lYYYr})
under the identification
$H=\e^{-Y}$.

Now let us consider the amplitudes on the disk $D^2$.
The non-trivial boundary operator
is the first cohomology class of the equator
$T$ of $\CP^1$
which is represented as the delta function 1-form at a point of $T$.
We denote it by $\vartheta$.
Let us count the number of deformations of maps in some classes.
First, the constant maps. Since the boundary $\partial D^2$ must be mapped
to the equator $T$, the whole disk must also be mapped to $T$.
Obviously, there is a one dimensional modulus --- the position in
$T$.
Next, degree one maps. As is well-known, $SL(2,\R)$ can be considered
as the parameter space of such maps
(it comes from the action
on the upper-half plane as $\zeta\to(a\zeta+b)/(c\zeta+d)$).
Thus, the parameter space is three-dimensional.
The maps of higher degree have more moduli.
For an amplitude to be non-vanishing,
the number of moduli must match with the axial R-charge of
the inserted operators.

Now let us consider an amplitude
with just a $\vartheta$ 
insertion at a point of the boundary $\partial D^2$.
Since one $\vartheta$ has axial R-charge $1$
only the constant maps can contribute.
The insertion of $\vartheta$ requires the insetion point
to be mapped to a given point in $T$.
There is only one such constant map.
Thus, we obtain
\beq
\langle \vartheta\rangle^{\pm}_{D^2}=1,
\label{lvtr}
\eeq
where the superscript $\pm$ distinguishes the $U(1)$ holonomy of the D-brane.
Next let us consider the case where $\vartheta$ is inserted
in $\partial D^2$ and $H$ is inserted in the interior of $D^2$.
The total axial R-charge is $3$ and thus, degree one maps
can contribute.
The insertion of $H$ requires the insertion point to be mapped to
a given point in $\CP^1$. This reduces the 3 moduli to one.
The insertion of $\vartheta$ further reduces the moduli
and there is only one map obeying the requirement.
The classical weight is $\e^{-t/2}$ or $\e^{-t/2+\pi i}$
depending on the Wilson line of the D-brane.
Thus. we obtain
\beq
\langle \vartheta H\rangle^{\pm}_{D^2}=\pm \e^{-t/2}.
\label{lvtttr}
\eeq
The results (\ref{lvtr}) and (\ref{lvtttr}) reproduce
the sine-Gordon result (\ref{lvtr}) and (\ref{lvtttr}).

\noindent
{\bf Remark.}
Open topological field theory has been studied from axiomatic point
of view in \cite{Moore,Laza1}.

\section*{Acknowledgement}

I would like to thank M. Douglas,
T. Eguchi, M. Gutperle, A. Iqbal,
H. Itoyama, S. Kachru, S. Katz, A. Lawrence, H. Liu,
J. McGreevy, G. Moore, M. Naka, M. Nozaki, B. Pioline, R. Thomas,
N. Warner, S.-K. Yang, E. Zaslow,
and especially E. Martinec for valuable
discussions.
I also thank M. Aganagic and C. Vafa for
explaining their work and for useful discussions.
I am grateful to KIAS, Seoul;
Aspen Center for Physics, Colorado;
SI-2000, Yamanashi;
and New High Energy Theory Center at Rutgers University
where parts of this work were carried out, for their hospitality. 
This work is supported in part by NSF-DMS 9709694.

\end{document}